\documentclass[twocolumn]{aastex62}

\usepackage{ulem}
\usepackage{url}
\usepackage{threeparttable}
\usepackage{xcolor}
\usepackage{soul}
\usepackage{booktabs}
\usepackage{natbib}
\usepackage{amsmath}
\usepackage{mathrsfs}
\hypersetup{colorlinks, linkcolor=black, anchorcolor=green, citecolor=blue}
\usepackage{txfonts}

\usepackage{rotating}

\usepackage[]{subfigure,graphicx}

\usepackage{multirow}

\received{}
\revised{}
\accepted{2020 Apr 7}
\submitjournal{ApJ}

\shorttitle{Chemical segregation of $\rm SO_2$ and SO toward the low-mass protostellar shocked region of L1157}
\shortauthors{Feng et al.}

\begin{document}

\title{Seeds of Life in Space ({\it SOLIS}). IX. Chemical segregation of $\rm SO_2$ and SO toward the low-mass protostellar shocked region of L1157Ê\footnote{} }

\correspondingauthor{Siyi Feng}
\email{siyi.s.feng@gmail.com}

\author[0000-0002-4707-8409]{S. Feng}
\affil{National Astronomical Observatory of China, Datun Road 20, Chaoyang, Beijing, 100012, P. R. China}
\affil{CAS Key Laboratory of FAST, NAOC, Chinese Academy of Sciences, P. R. China}
\affil{Max-Planck-Institut f\"ur Extraterrestrische Physik, Gie{\ss}enbachstra{\ss}e 1,  D-85748,  Garching bei M\"unchen, Germany}

\author{C. Codella}
\affil{INAF-Osservatorio Astrofisico di Arcetri, Largo E. Fermi 5, I-50125, Florence, Italy} 
\affil{Univ. Grenoble Alpes, CNRS, IPAG, F-38000 Grenoble, France} 

\author{C. Ceccarelli}
\affil{Univ. Grenoble Alpes, CNRS, IPAG, F-38000 Grenoble, France} 

\author{P. Caselli}
\affil{Max-Planck-Institut f\"ur Extraterrestrische Physik, Gie{\ss}enbachstra{\ss}e 1,  D-85748,  Garching bei M\"unchen, Germany}

\author{A. Lopez-Sepulcre}
\affil{Univ. Grenoble Alpes, CNRS, IPAG, F-38000 Grenoble, France} 
\affil{Institut de Radioastronomie Millim\'etrique, 300 rue de la Piscine, Domaine Universitaire de Grenoble, F-38406 Saint-Martin d'H\`eres, France}

\author{R. Neri}
\affil{Institut de Radioastronomie Millim\'etrique, 300 rue de la Piscine, Domaine Universitaire de Grenoble,  F-38406 Saint-Martin d'H\`eres, France} 

\author{F. Fontani}
\affil{Max-Planck-Institut f\"ur Extraterrestrische Physik, Gie{\ss}enbachstra{\ss}e 1,  D-85748,  Garching bei M\"unchen, Germany}
\affil{INAF-Osservatorio Astrofisico di Arcetri, Largo E. Fermi 5, I-50125, Florence, Italy}

\author{L. Podio}
\affil{INAF-Osservatorio Astrofisico di Arcetri, Largo E. Fermi 5, I-50125, Florence, Italy} 

\author{B. Lefloch}
\affil{Univ. Grenoble Alpes, CNRS, IPAG, F-38000 Grenoble, France} 

\author{H. B. Liu}
\affil{Academia Sinica Institute of Astronomy and Astrophysics, No.1, Sec. 4, Roosevelt Road, Taipei 10617, Taiwan, Republic of China} 

\author{R. Bachiller}
\affil{Observatorio Astronomico Nacional (OAN, IGN), Calle Alfonso XII, 3, E-28014 Madrid, Spain}

\author{S. Viti}
\affil{ Department of Physics and Astronomy, University College London, Gower Street, London WC1E 6BT, UK}



\begin{abstract} 
We present observations of SO and $\rm SO_2$ lines toward the shocked regions along the L1157 chemically rich outflow, taken in the context of the Seeds Of Life In Space IRAM-NOrthern Extended Millimeter Array Large Program, and
supported by data from Submillimeter Array and  IRAM-30\,m telescope at 1.1--3.6\,mm wavelengths.  We simultaneously analyze, for the first time, all of the brightest shocks in the blueshifted lobe, namely, B0, B1, and B2. 
We found the following.
(1) SO and $\rm SO_2$ may trace different gas,  given 
that the large(-scale) velocity gradient analysis indicates for $\rm SO_2$  a volume density ($\rm 10^5\text{--}10^6\,cm^{-3}$) denser than that of the gas emitting in SO by a factor up to an order of magnitude.
(2) Investigating the 0.1\,pc scale field of view, we note a tentative gradient along the path of the precessing jet.
More specifically, $\rm \chi({SO/SO_2})$ decreases from the B0-B1 shocks to the older B2.
(3) At a linear resolution of 500--1400\,au, a tentative spatial displacement between the two  emitting molecules is detected, with the SO peak closer (with respect to $\rm SO_2$)  to the position where the recent jet is impinging on the B1 cavity wall. 
Our astrochemical modeling shows that the SO and $\rm SO_2$ abundances
evolve on timescales less than about 1000\,years.
Furthermore, the modeling requires high 
abundances ($2\times10^{-6}$)  of both $\rm H_2S/H$ and S/H injected in the gas phase due to the shock occurrence,
so pre-frozen OCS only is not enough to reproduce our new observations.
\end{abstract}


\section{Introduction}
 
In the earliest stage of star formation, supersonic jets generated by  protostars  impact on the natal dense parent cloud and create shocks \citep[e.g.,][]{frank14}. Shocks compress and heat the surrounding gas, driving complex chemistry in a short timescale, including endothermic chemical reactions, ice mantle sublimation, and sputtering. These chemical processes are infeasible toward the pre-shock gas \citep[e.g.][]{viti11}.
As a result, gas-phase abundances of species such as sulfur (S)-bearing molecules are enhanced by several orders of magnitude compared with the prestellar cores and protostellar envelopes \citep[e.g.][]{tieftrunk94,kaufman96,bachiller01,jimenez05,jorgensen07}.

\subsection{S-bearing chemistry}
S-bearing molecules are highly reactive in the gas phase and are very sensitive to the thermal and kinetic properties of the gas. 
It has long been proposed that the relative abundance ratio between  $\rm H_2S$, $\rm SO_2$, and SO can be used to date the evolutionary stages of star-formation (i.e., constraining the age of the outflow and associated shocks; \citealp{charnley97,hatchell98,buckle03,wakelam04}). For example, in agreement with theoretic predictions, observations revealed that the $\rm SO_2/SO$ ratio is enhanced in young  ($\rm <10^3$\,yr) shocked region compared with the cloud, due to gaseous endothermic reactions; 
while this ratio decreases for the conversion of $\rm SO_2$ to atomic sulfur in the old shocks \citep[$\rm >3\times10^4$\,yr, e.g.,][]{pineau93,hatchell98,wakelam04,feng15}.
However, the feasibility of taking the relative abundances of  S-bearing species as a star formation dating tool depends on the physical condition of the source and a proper assumption of the S grain reservoir. 

{\color{black} Moreover, understanding of the S-bearing chemistry is challenged by a collection of mysteries, the major one being the ``S-depletion problem" \citep[e.g, ][]{tieftrunk94}; i.e., although S is known to be significantly depleted in molecular dark clouds, its depleted form (the main reservoir on grain mantles) is far from being established.}
Several species have been proposed to be the main carriers of S on dust grains, such as $\rm H_2S$ \citep[e.g.,][]{lij15,holdship16}  and OCS \citep[e.g.,][]{hatchell98,wakelam04,liu12,podio14,minh16,holdship19}. Historically, $\rm H_2S$ has never been definitively detected on ices, which calls for further and more comprehensive S-chemistry study using telescopes at higher sensitivity. Recent observations and models have shown that, at the very least, $\rm H_2S$ is not the only S-bearing species on grains \citep[e.g.,][]{wakelam04b,codella05}. By far, OCS is detected \citep[e.g.,][]{geballe85,palumbo95,podio14,boogert15}, and $\rm SO_2$ \citep[e.g.,][]{boogert97} is tentatively detected in interstellar ices. Therefore, the main reservoir of S on dust is still unclear.

\subsection{The blueshifted outflow lobe of L1157}
Regions containing successive young ($\lesssim$ 10$^3$ yr) shocks driven by mass losses from protostars are ideal laboratories for studying  time-dependent S-chemistry, as the short time-scales of the shocks provide the possibility to either directly observe the material coming off grains, or to trace the chemistry back to the original content of interstellar ices.

This well applies to the target of the present study, the shocks associated with the L1157 outflow region.
The Class 0 protostar L1157-mm \citep[$\simeq$ 3 $L_{\rm \odot}$,][]
{tobin10}, located at $d\sim$352\,pc \citep{zucker19}, drives an episodic and precessing jet \citep[e.g.][]{gueth96,gueth98,podio16}, 
creating a bipolar outflow \citep[e.g.,][]{bachiller01,nisini10}.
The impacts between the jet and the high-density \citep[$\rm 10^4\text{--}10^5\,cm^{-3}$, ][]{lefloch12,benedettini13,gomezruiz15} cavity walls triggered several shocks in the last $\rm \sim2\times 10^3$\,yr \citep[e.g.,][]{lefloch10}.

The targets of the present work are the B0, B1, and B2 shocked regions, which are the brightest shocks
associated with the southern blueshifted outflow lobe, 
 located within 0.1\,pc from the protostar in the plane of the sky.
Specifically, the B1 region has been studied in detail by using either single-dish telescopes (e.g., {\it Herschel-CHESS}, \citealt{ceccarelli10}; {\it 30\,m-ASAI}, \citealt{lefloch18}) or interferometers \citep[SMA, Very Large Array, NOEMA,][]{tafalla95,codella09,lefloch12,benedettini13}, revealing a clumpy bow-like structure associated with several chemical processes, such as the sputtering of dust mantles, the disruption of the dust refractory cores, and warm gas-chemistry reactions.
The abundance of molecular species,
from diatomic--such as CO, CS--to molecular ions--such as $\rm HOCO^+$, $\rm SO^+$, and $\rm HCS^+$ \citet{podio14}--and more complex molecules such as $\rm CH_3OH$, $\rm CH_3CHO$, and $\rm NH_2CHO$ \citep[e.g., ][]{bachiller01,benedettini07,arce08,codella10,codella15,codella17,busquet13,fontani14c,lefloch17,lefloch18}  are consequently enhanced.
Specifically, S-bearing species (e.g., CS, SO, $\rm SO_2$, $\rm H_2S$, OCS) have been detected toward this region  \citep[see, e.g.,][]{bachiller01,benedettini13,podio14,holdship16}.

Kinematically, B2 is older than B1 \citep[e.g., ][]{bachiller01,podio16}; B1 mainly consists of three shock layers produced by episodes of the precessing jet, with the oldest layer located the farthest from the protostar, i.e. in the southernmost edge of B1  \citep{codella15}.
Observing the jet radial velocity in the inner knots, the precessing model \citep[][]{podio16} indicates that\footnote{\citet[][]{podio16} estimated the kinematic ages of each knot by using  $d\sim$250\,pc. The kinematic ages of these knots in this work are updated by using $d\sim$352\,pc \citep{zucker19}.} B0 is $\sim$1340\,yr old,  B2 is  $\sim$2530\,yr old, and the kinematic age of B1 knots ranges from 1550\,yr at the north edge  to 1760\,yr at the southern edge.
Therefore, the entire B0-B1-B2 region provides us one of the best laboratories to study the time-dependent S-chemistry.

In this paper, we present the data in Section~\ref{1157:obs}, which are obtained from the ``Seeds Of Life In Space"\citep[{\it SOLIS\footnote{The ``Seeds Of Life In Space" large programme is aiming at investigating molecular formation and processes during the early stages of the
star formation process, by using IRAM NOrthern Extended Millimeter Array (NOEMA).};}][]{ceccarelli17} large program and supplementary Submillimeter Array (SMA) observations. Our results are presented in Section~\ref{1157:result} and analyzed in Section~\ref{1157:analysis}. We study the map of relative abundance ratio between SO and $\rm SO_2$  in Section~\ref{1157:ratio}, and 
discuss the possible origin of SO and $\rm SO_2$ by using our chemical model to fit the observation measurement in Section~\ref{1157:model}.
 We conclude in Section~\ref{1157:conclusion}.

\section{Observations}\label{1157:obs}
Using the NOrthern Extended Millimeter Array (NOEMA) and the SMA, we carried out a line-imaging survey at 1.1--3.6\,mm toward our targets.
Given that the antennae of NOEMA and SMA have different diameters, and that the molecular lines we observed are at different wavelengths, the primary beam attenuation is different. Therefore, the B0, B1, and B2 shocked regions have been imaged by using one point or two  mosaic points with NOEMA or SMA at different wavelengths.

\subsection{NOEMA}
Using NOEMA, we observed B0-B1 at 3.6\,mm, 3.1\,mm, and 1.5\,mm in its A (7 antennae), C  (6-8 antennae), and D (5-8 antennae) configurations from 2015 July to 2017 January. With the shortest projected baseline of 24\,m, the observations are sensitive to structures up to 19\arcsec\,(at 3.6\,mm), 16\arcsec\,(at 3.1\,mm), and 8\arcsec\,(at 1.5\,mm), respectively.
For all  observations, a common phase center {at $\rm 20^h39^m10^s.20$, $\rm 68^{\circ}01^{'}10^{''}.50$ (J2000) and  the systemic  velocity  $\rm V_{sys}\sim 2.7\,km\,s^{-1}$ was used. The precipitable water vapor  (PWV) varied between 1 and 40 mm during the observations.
Standard interferometric calibrations were performed during the observations. The bandpass,  gain (phase/amplitude), and flux calibrators used during the observations are recorded in Table~\ref{NOEMAconfI}. The uncertainty of absolute flux scale is estimated to be correct to within $\rm \sim 5\%$.

With the  wide-band receiver (WIDEX), three spectral setups cover the frequency ranges of 80.790--84.407, 95.794-99.499, and 203.962-207.632\,GHz,  at a frequency resolution of 1.95\,MHz. We detected one $\rm SO_2$ line and three SO ($\rm ^{34}SO$) lines (with $E_u/k_B \rm<40\,K$).
From our {\it ASAI}-30\,m single-point observations at a frequency resolution of 0.195\,MHz  \citep{lefloch18}, these lines have full-width half-maximum (FWHM) linewidth of $\rm>8\,km\,s^{-1}$ toward the B1 shocked region \citep{mendoza14}, so the velocity resolution from WIDEX observations ($\rm \sim7\,km\,s^{-1}$ at 83.688\,GHz, see Table~\ref{line} and Figure~\ref{1157:lpeg}) is sufficient for our chemical analysis. 
Moreover, we use the narrow-band correlator  and set four windows in each spectral setup 
 (80\,MHz backends with a spectral resolution of 156\,kHz) to target on several complex organics lines \citep[see the spectral setups in][for details]{ceccarelli17,codella17}. Other lines detected toward this region will be presented in the follow-up
papers (e.g., CS/CCS/OCS/NS/SiS/$\rm H_2S$/$\rm H_2CS$, cyanopolyynes; S. Feng et al. in prep.).

Furthermore, aiming at a robust analysis on the possible chemical differentiations between $\rm SO_2$ and SO toward different time-dependent shock layers of B1, we also include one $\rm ^{34}SO$ and four $\rm SO_2$ ($\rm ^{34}SO_2$) lines from our previous NOEMA-WIDEX observations (\citealp[obsID: W068,][]{fontani14c}; \citealp[obsID: X058,][]{podio17}; see observation details therein). The spectroscopic parameters of all these lines are taken from the Cologne Database for Molecular Spectroscopy \citep[CDMS,][]{mueller02,muller05} and listed in Table~\ref{line}.

Calibration and imaging were performed by using the standard procedures with the CLIC and MAPPING softwares of the GILDAS\footnote{http://www.iram.fr/IRAMFR/GILDAS} package.
After testing different weighting algorithms, we used the natural  weighting to achieve a better signal-to-noise ratio (S/N).  However,  our NOEMA data recover only  40--60\% of the emission when compared with our {\it ASAI}-30\,m pointing observations, indicating that these $\rm SO_2$ and SO lines should have extended emissions toward our target.

{\color{black}Averaging the line-free channels of the WIDEX band, we did not detect significant ($\rm>5\sigma$ rms) continuum emission at 3.6\,mm with a synthesized beam of $\rm 5.08^{''}\times 4.69^{''}$ ($\rm \sigma$=$\rm 0.016\,mJy\,beam^{-1}$), at 3.1\,mm with a synthesized beam of   $\rm 3.12^{''}\times 2.30^{''}$ ($\rm \sigma$=$\rm 0.012\,mJy\,beam^{-1}$), or 1.5\,mm with a synthesized beam of   $\rm 1.20^{''}\times 0.85^{''}$ ($\rm \sigma$=$\rm 0.036\,mJy\,beam^{-1}$).
}

 From our NOEMA observations, the detected SO isotopologue lines have low-$J$ levels ($E_u/k_B \rm$=9--18\,K), while the $\rm SO_2$ isotopologue lines have mid-$J$ or high-$J$ levels ($E_u/k_B \rm$=16--102\,K). They may selectively trace gas with different temperatures and therefore show different spatial distributions. With the aim of investigating any chemical difference from the spatial distribution maps of both species, it is necessary to observe more $\rm SO_2$ and SO lines, to cover the same range of  upper energy level  ($E_u/k_B$).

\subsection{SMA}
To minimize the excitation effect on the  spatial distribution differentiation between  SO and $\rm SO_2$, we carried out four-track observations toward our source with the SMA at 230/255\,GHz (1.14-1.40\,mm), by using the compact configuration. From 2017 June to July, we observed B1 and B2 with a two-point mosaic per track.  The phase centers of B1 and B2 are at $\rm 20^h39^m09^s.500$, $\rm 68^{\circ}01^{'}10^{''}.500$ and $\rm 20^h39^m11^s.358$, $\rm 68^{\circ}00^{'}51^{''}.00$ (J2000), respectively. 
For all of  the observations, the phase and amplitude calibrations were performed via frequent switch (every 20 minutes) on quasars 1849+670 and 1927+739. 

The baselines range from 16\,m to 76\,m with 7 or 8 antennae at different dates. B1 and B2 share the bandpass (3C273 or 3C454.3) and flux  (Callisto and MWC349a)  calibrators.
The zenith opacities,  measured with water vapor monitors mounted on the James Clerk Maxwell Telescope (JCMT),  were satisfactory during all tracks with $\tau (\rm 225~ GHz)\sim0.05\text{--}0.3$.
The system temperature was 100--200\,K at 230\,GHz and 200--400\,K at 255\,GHz.
Further technical descriptions of the SMA and its calibration schemes can be found in \citet{ho04}.

We employed the dual-RX, single polarization and double-sideband observing mode. Our frequency coverages (214.935--221.030, 229.004--235.015, 240.738--246.828, 254.804--260.890\,GHz) were sampled using the SWARM correlator with a 0.140\,MHz frequency resolution ($\sim$0.182\,km\,s$^{-1}$ at 230.538\,GHz). We tuned 230.0\,GHz and 255.8\,GHz to be at the center of the spectral chunk 1 in the upper sideband of the receivers Rx230 and Rx240, respectively.
To increase the S/N ratio per channel, we smooth the frequency resolution to 4.469\,MHz  (a velocity resolution of 5.812$\rm ~km\,s^{-1}$ at 230.538\,GHz).

The  flagging and calibration was done with the MIR package \citep{scoville93,qi03b}\footnote{The MIR package  was originally developed for the Owens Valley Radio Observatory, and is now adapted for the SMA, http://cfa-www.harvard.edu/$\sim$cqi/mircook.html. }.  The  mosaic imaging was conducted with MIRIAD package by using the natural  weighting \citep{sault95}. 

We detected four SO lines in the entire frequency coverage ($E_u/k_B \rm$=35--56\,K), while no $\rm SO_2$ line is detected with $\rm >3\sigma$ emission {($\rm \sigma$=$\rm 42.7\,mJy\,beam^{-1}$ at an angular resolution of $\rm \sim3.4^{''}$)}. Their spectroscopic parameters taken from CDMS are listed in Table~\ref{line} as well. In addition, we also detected lines from SiO, HNCO, CO, CS, $\rm H_2CO$ in these observations.

Averaging the line-free channels, the sensitivity of continuum emission  at 1.1--1.4\,mm achieves $\rm \sigma$=$\rm 0.17\,mJy\,beam^{-1}$ with a synthesized beam of $\rm 3.37^{''}\times 3.13^{''}$.  At such sensitivity, we have marginally detected 3--4\,$\sigma$ of continuum emission  toward B1-B2 region (see H. B. Liu in prep. for a detailed continuum study).

\subsection{IRAM-30\,m Telescope}
{To compensate the missing flux, we use IRAM-30\,m telescope and performed the observations in the on-the-fly (OTF) mode from 2017 May to November. We mapped a $2.5'\times2.5'$ area (at least twice that of the primary beam of the interferometers for successful combination, according to \citealt{pety13}, covering both B1 and B2), centered at $\rm 20^h39^m09^s.50$, $\rm 68^{\circ}01^{'}10^{''}.5$ (J2000). 
By superpositioning different spectral tunings, the broad bandpass of EMIR (16\,GHz bandwidth simultaneously for each spectral tuning)  covers the frequency range of 
79.302--87.084, 91.821--102.764, 106.902--114.683, 130.703--138.485, 161.102--169.882, 203.925--211.703,
213.422--227.385, 229.105--236.882, 239.420--247.204, and 255.103--262.880
\,GHz, with a frequency resolution of 0.195\,MHz (by using the FTS200 backend).
The FWHM beam of the 30\,m telescope at 83.688\,GHz is $\sim30.86\arcsec$.

To combine the  emissions of $\rm SO_2$ and SO isotopologue lines obtained from NOEMA or SMA at high spatial resolution with their large-scale emission data obtained from IRAM-30\,m, we aligned the phase center and smoothed the velocity resolution of lines obtained from IRAM-30\,m  to  those from NOEMA or SMA. Afterwards, we use the task ``UV-SHORT" in the MAPPING software for NOEMA-30\,m combination; 
and use the MIRIAD package for SMA-30\,m combination. 
The synthesized beam and the $\rm 1\sigma$ rms value of the data cube of each line, after NOEMA-30\,m or SMA-30\,m combination, are listed in Table~\ref{line}.

\begin{table*}
\small
\centering
\caption{Target lines in this work with their spectroscopic and observation parameters
 \label{line}}
 \scalebox{0.92}{
\begin{tabular}{p{0.8cm}ccccccccc}
\hline\hline
Mol. &Freq.$^a$  &Transition$^a$  &$\rm S\mu^2$  
&$E_u/k_B \rm$$^a$  &$\rm \Delta V$   &$\rm \Delta \theta$ (P.A.) $^b$  &rms$^b$   &Primary beam  &Interferometric\\
         &(GHz) &                                                &($\rm D^2$)   
          &  (K)              &($\rm km\,s^{-1}$)                                &($\rm \arcsec\times \arcsec$, $^\circ$)                                         &($\rm mJy\,beam^{-1}\,ch^{-1}$)              &(\arcsec)     &Observations\\    
\hline

$\rm SO_2$    &	83.688           &$\rm 8_{1,7}-8_{0,8}$    &17.01 
&37  &7.01         &$\rm 5.30\arcsec\times4.86\arcsec~(-94^\circ)$            &0.7               &$\rm 61$     &{\it SOLIS}-NOEMA\\  
$\rm SO_2$    &134.004         &$\rm 8_{2,6}-8_{1,7}$   &15.18   
&43  &4.37         &$\rm 2.64\arcsec\times2.43\arcsec~(-89^\circ)$           &1.1           &$\rm 38$       &NOEMA$^{c}$\\ 
$\rm SO_2$    &135.696          &$\rm 5_{1,5}-4_{0,4}$   	&8.35 	
&16  & 4.32        &$\rm 2.65\arcsec\times2.51\arcsec~(75^\circ)$            &1.3             &$\rm 38$       &NOEMA$^{c}$\\  
$\rm SO_2$    &163.606         &$\rm 14_{1,13}-14_{0,14}$   &17.10  	
&102  & 3.58        &$\rm 2.01\arcsec\times1.87\arcsec~(43^\circ)$            &1.4          &$\rm 31$          &NOEMA$^{d}$\\  
$\rm ^{34}SO_2$    &	162.776          &$\rm 7_{1,7}-6_{0,6}$    &11.86 
&26  &3.60        &$\rm 1.98\arcsec\times1.86\arcsec~(47^\circ)$            &0.9               &$\rm 31$     &NOEMA$^{d}$\\  

$\rm SO$    &99.230         &$\rm 3_{2}-2_{1}$     &6.91 
&9  &5.90         &$\rm 2.95\arcsec\times2.20\arcsec~(25^\circ)$            &0.4              &$\rm 52$     &{\it SOLIS}-NOEMA\\  
$\rm SO$    &206.176       &$\rm 4_{5}-3_{4}$    &8.91 	
&39    &2.84         &$\rm 1.15\arcsec\times0.80\arcsec~(-171^\circ)$            &1.3         &$\rm 25$           &{\it SOLIS}-NOEMA\\  
$\rm SO$   &215.221               &$\rm 5_{5}-4_{4}$    &11.31   
&44  &6.23 &$\rm 3.91\arcsec\times3.66\arcsec~(-2^\circ)$   &21.9 &$\rm 55$     &SMA\\
$\rm SO$   &219.949               &$\rm 6_{5}-5_{4}$     &14.01  	
 	&35  &6.09 &$\rm 3.82\arcsec\times3.58\arcsec~(-3^\circ)$   &32.1 &$\rm 53$     &SMA\\
$\rm SO$   &258.256             &$\rm 6_{6}-5_{5}$    &13.74   	
&56  &5.19 &$\rm 3.23\arcsec\times3.04\arcsec~(-4^\circ)$   &22.2 &$\rm 45$     &SMA\\
$\rm SO$   &261.844              &$\rm 7_{6}-6_{5}$    &16.38   
	&48  &5.12 &$\rm 3.18\arcsec\times3.00\arcsec~(-4^\circ)$  &73.1 &$\rm 45$     &SMA\\
$\rm ^{34}SO$    &97.715      &$\rm 3_{2}-2_{1}$  &6.92 
&9  &5.99         &$\rm 2.98\arcsec\times2.33\arcsec~(25^\circ)$            &0.4          &$\rm 52$          &{\it SOLIS}-NOEMA\\  
$\rm ^{34}SO$    &135.776         &$\rm 4_{3}-3_{2}$   &9.28  	
&16  &4.31         &$\rm 2.64\arcsec\times2.50\arcsec~(75^\circ)$            &1.2       & $\rm 38$            &NOEMA$^{c}$\\ 
$\rm SiO$    &217.105         &$\rm 5-4$   &48.0  	
&32  &6.17         &$\rm 3.99\arcsec\times3.75\arcsec~(-3^\circ)$            &22.7       & $\rm 54$            &SMA\\ 

\hline
\hline
\multicolumn{10}{l}{{\bf Note.} {\it a}.  Spectroscopic parameters are taken from the CDMS.}\\
\multicolumn{10}{l}{{~~~~~~~~~} {\it b.}  Data from NOEMA-30\,m or SMA-30\,m combination, without angular resolution smoothing,}\\
\multicolumn{10}{l}{{~~~~~~~~~~~~~} measured with the referred beam size (``beam") and channel width (``ch"). }\\
\multicolumn{10}{l}{{~~~~~~~~~} {\it c}. Unpublished data was taken from NOEMA observations W068 reported by \citet{fontani14c}.}\\  
\multicolumn{10}{l}{{~~~~~~~~~} {\it d}.  Unpublished data was taken from NOEMA observations X058 reported by \citet{podio17}.}\\
\end{tabular}
}

\end{table*}


\section{Molecular spatial distribution}\label{1157:result}
 For the first time, the interferometric data of L1157 B0-B1-B2 are complemented  with the IRAM-30\,m mapping, providing us images with full spatial scale coverage, as well as high angular resolutions. Moreover, the wide frequency coverage allows us to study five $\rm SO_2$ lines (including one $\rm ^{34}SO_2$ line) and  eight SO lines  (including two $\rm ^{34}SO$ line), with $E_u/k_B \rm$ in the range of 16--102\,K.
 At a linear resolution of $\rm<1000$\,au, we will be able to search for  chemical differentiations from the systematic differentiations in the spatial distribution of both species.

 \begin{figure}
\centering
\includegraphics[width=6cm]{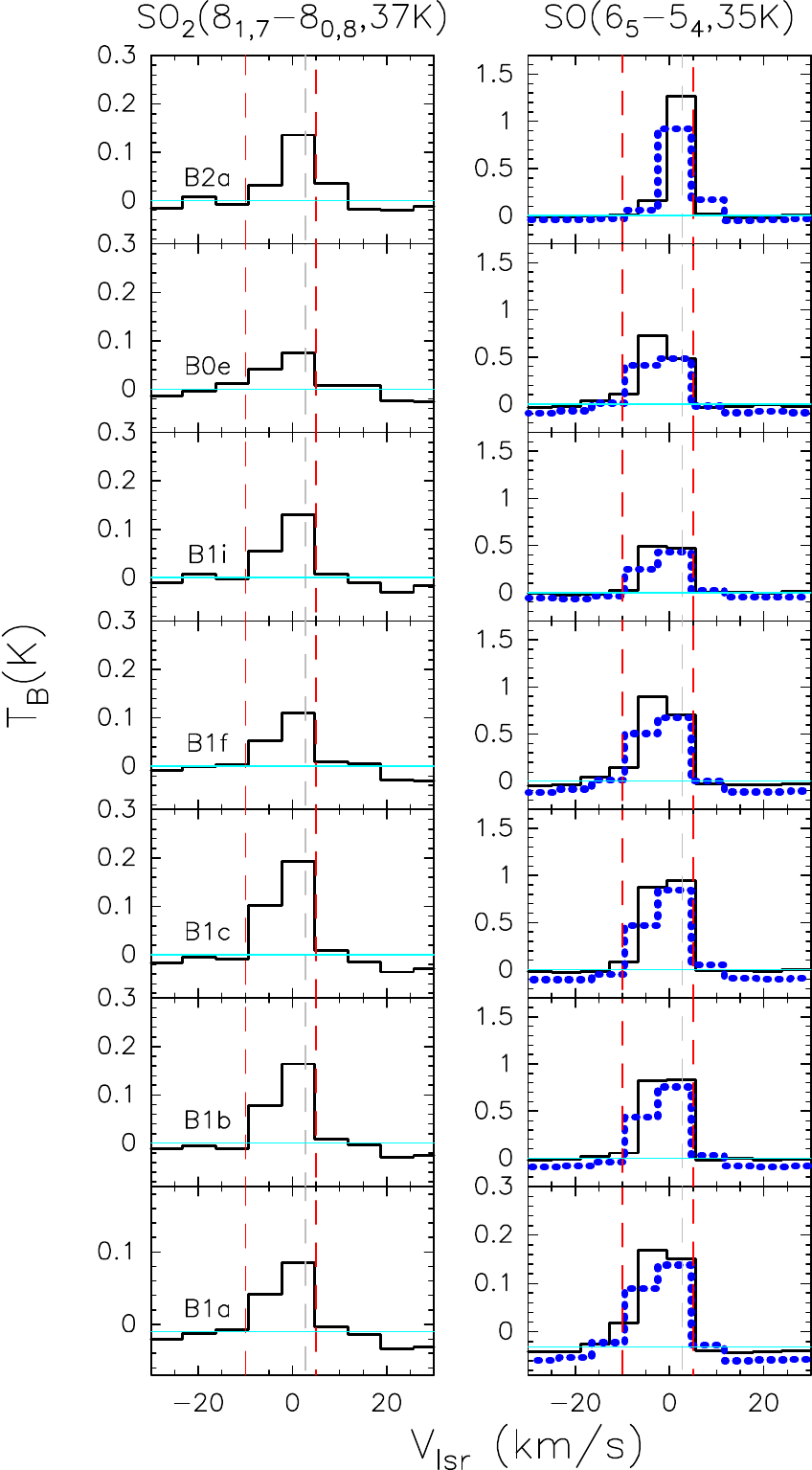}
\caption{
Profiles of two identified SO and $\rm SO_2$ lines with the same $E_u/k_B$ and similar beam attenuation.  The profiles of all identified SO and $\rm SO_2$ lines are shown in Appendix Figure~\ref{1157:velpro}.
Lines are averaged from a beam-sized region with the center  toward each clumpy substructure in the plane of the sky  (Table~ \ref{tab:clump}). 
 All lines are extracted from images smoothed to the same pixel size and the same angular resolution ($\rm 5.1\arcsec$). The  line profile of SO\,($\rm 6_5\text{--}5_4$) shown in black or dashed blue has a velocity resolution of 6.09\,$\rm km\,s^{-1}$ (original spectral resolution) or 7.01\,$\rm km\,s^{-1}$ (the same as that of $\rm SO_2\,8_{1,7}\text{--}8_{0,8}$), respectively. In each panel, two red vertical lines at $\rm -10.0\,km\,s^{-1}$ and $\rm +5.0\,km\,s^{-1}$ indicate the velocity range for which we integrate the intensity; the gray dashed vertical line indicate the $\rm V_{sys}=2.7\,km\,s^{-1} $ of the cloud. The horizontal cyan line indicates the baseline ($\rm T_B$=0\,K).}\label{1157:lpeg}
\end{figure}

 All lines show a significant blueshifted emission toward B0-B1-B2,  and the $\rm S/N>3$ part of each line is in the velocity range of $\rm -10$ to $\rm +5\,km\,s^{-1}$  (Figure~\ref{1157:lpeg} and ~\ref{1157:velpro}). 
Specifically, the peaks of all $\rm SO_2$ lines are close to the $\rm V_{sys}$, when observing at a velocity resolution of 3.6--7.0\,$\rm km\,s^{-1}$. 
Observed at higher velocity resolutions (2.8--5.9\,$\rm km\,s^{-1}$), some SO lines seem to  peak at $\rm \sim-2\,km\,s^{-1}$ toward positions such as B0e, B1f, and B1a.  However, such a velocity offset  ($\rm \sim5\,km\,s^{-1}$) only corresponds to 1--2 channel(s), and it smears out when these SO lines are binned to a coarser velocity resolution (7.0\,$\rm km\,s^{-1}$). 
Therefore, whether  these SO and $\rm SO_2$ lines trace different gas calls for further observations at a higher-velocity resolution.

We imaged the spatial distribution of each line by integrating their emissions over the velocity range\footnote{We compared the integrated intensity of each line over the velocity range of $\rm -10$ to $\rm +5\,km\,s^{-1}$ ($\rm S/N>3$) with $\rm -20$ to $\rm +15\,km\,s^{-1}$ (line emission down to zero), and found the difference between these integrations is $\rm <5\%$, which is less than the systematic uncertainty from observations and data reduction.} of $\rm -10$ to $\rm +5\,km\,s^{-1}$. 
Specifically, Figure~\ref{1157:jet} show the spatial distributions of $\rm SO_2\,(8_{1,7}-8_{0,8})$ and  four SO  lines, i.e., $\rm 6_{5}-5_{4}$, $\rm 5_{5}-4_{4}$, $\rm 7_{6}-6_{5}$, and $\rm 6_{6}-5_{5}$, which are sensitive to a comparably larger field than the rest of the lines (see their primary beams in Table~\ref{line}). Overlaid with CO\,(1--0) emission contours \citep{gueth96}, we found that all of these  SO and $\rm SO_2$ line emissions  trace the extended structure from  the eastern cavity wall B0 to two cavities associated with B1 and B2 \citep[denoted as C2 and C1, respectively in ][]{gueth98}.
These lines show the same shock knots in B0 and B1 as those in SiO\,(5--4)  from the same SMA-30\,m observations\footnote{The combination of two-point mosaic of SMA observations and IRAM-30\,m observations recover extended emission of SiO\,(5--4), which that was missed toward B2 in previous  SMA-only observations \citep{gomezruiz13}.} (Figure~\ref{1157:jet}) as well as SiO\,(2--1) observed by NOEMA-30\,m \citep{gueth98},  which is consistent with the prediction of a precessing jet model \citep{podio16}.

 \begin{figure*}
\centering
\includegraphics[width=18cm]{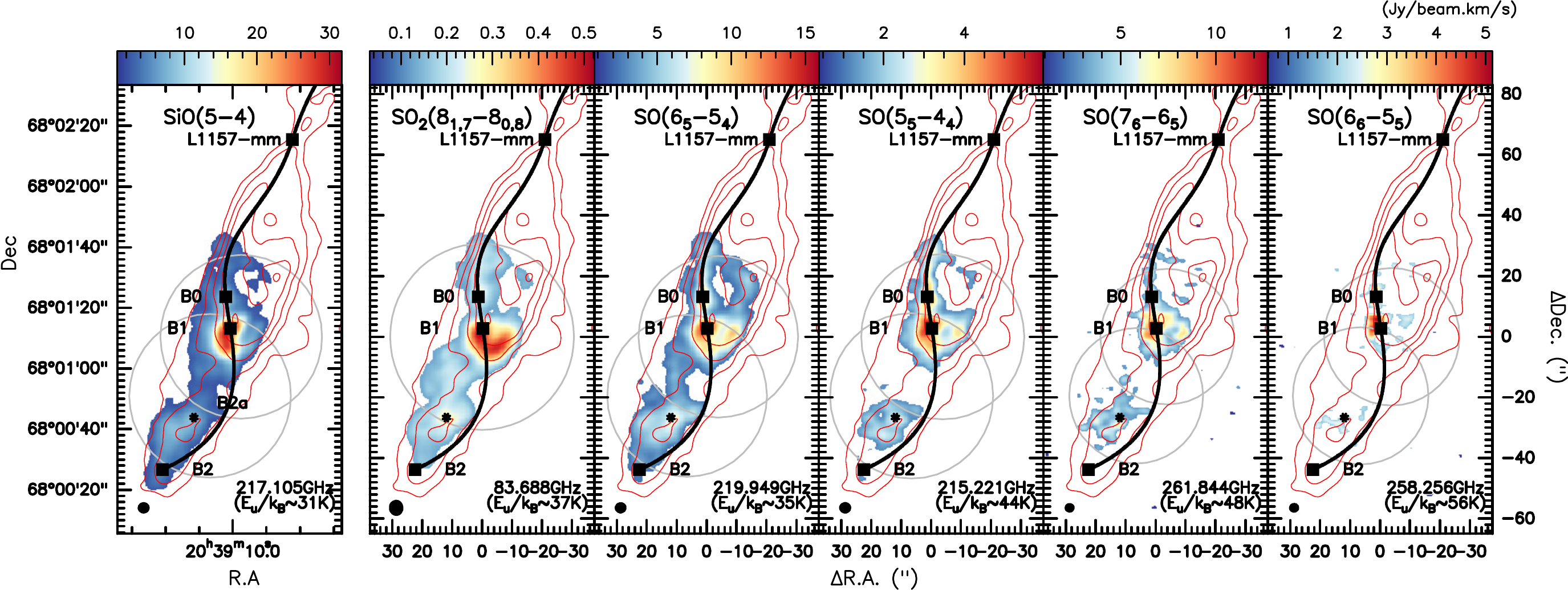}
\caption{
Outline of the southern (blueshifted) outflow lobe from the low-mass protostar L1157-mm. Color maps show the intensity maps of several shock tracer lines integrated over the velocity range of $\rm -10$ to $\rm +5\,km\,s^{-1}$ , which are sensitive to the extended B0-B1-B2  region. 
All of the maps are obtained from the combination of NOEMA (or SMA) and IRAM-30\,m, and line emissions with $\rm <5\sigma$ rms are blanked. The red contours, starting from $\rm 8\sigma$ ($\rm \sigma=0.27\,Jy\,beam^{-1}$) and increasing with the step of  $\rm 8\sigma$, show the CO (1-0) emission, 
obtained from NOEMA and IRAM-30\,m combination \citep{gueth96}.  Gray circle(s) in each panel indicates the primary beam of the NOEMA/SMA at corresponding line rest frequency, and the  angular resolution  after IRAM-30\,m combination is plotted at the bottom left in each panel.
The black curve guidelines the path of the precessing jet from L1157-mm as modeled by \citet[][]{podio16}. The black squares indicate the driving source, L1157-mm, and the shock knots in  B0, B1, and B2.  }
\label{1157:jet}
\end{figure*}

Aside from the above five lines, the rest SO and $\rm SO_2$ isotopologue lines in this work are only sensitive to the B0-B1 region.  Nevertheless, they offer images observed at higher angular resolutions.
At a spatial resolution of 500--1400\,au,  
six clumpy substructures appear toward the B0-B1 region  in Figure~\ref{1157:smalldistribution}. 
These substructures are already identified from previous observations {   \citep[e.g., by using line emissions from $\rm CH_3CN$, HCN, $\rm CH_3OH$, SiO, CS, $\rm H_2CO$, $\rm HCO^+$, and HNCO in ][]{benedettini07,codella09,gomezruiz13,burkhardt16}.
 In this work, we use the consistent names of  B0e, B1a, {  B1b},  {  B1c},  and B1f with those defined by \citet{benedettini07,codella09,burkhardt16}.
Moreover, we denote a position {  B1i}, which was noted previously but not labeled as a clumpy substructure from emission of other species. 
For any particular substructure traced by more than one molecule, the difference in the measurement of its absolute coordinate by different molecular tracers is about 1\arcsec--2\arcsec. Based on Figure~\ref{1157:smalldistribution}, the absolute coordinates of these clumpy substructures are listed  in Table~\ref{tab:clump}.

Morphologically, B0e outlines the eastern wall of the northernmost blueshifted bow-shock cavity \citep[e.g., named as ``C2" in][]{gueth98}, showing higher abundance of species such as SiO and HCN than the western wall and the connecting ridge B1f \citep{gomezruiz13} .
B0e and B1a  are suggested as knots where episodic ejection impacting against the cavity wall   \citep{podio16}, and high-velocity SiO  and CS emissions  \citep{benedettini07,gomezruiz13} are detected toward them.
A U-shape structure connecting B1a-B1c-B1b pins down the shock front, and the apex B1i is coincident with the ``finger" structure detected on the maps of CS and low-velocity SiO, which is suggested as the magnetic precursor of the shock  \citep{gueth98}.

According to the spatial distributions of detected complex organic molecules (COMs), several chemical layers are defined in B1.
One chemical layer is composed of B0e-B1a-B1f-B1b,  where the detected COMs include HDCO \citep[][]{fontani14}, $\rm CH_3CHO$ \citep[][]{codella17}, and $\rm CH_3OH$ \citep[][]{codella17}, indicating the place where the current dust {mantle and core are released} by sputtering. 
To the southern shock front, B1c and B1i denote another chemical layer where $\rm NH_2CHO$ is the main COM emitter \citep[][]{codella17}, and this layer contains the gas older than B1a-B1f-{  B1b} \citep{gueth98,podio16}. 

}

To compare the line emissions along the path of the processing jet, we also denote a position {  B2a}, where SO and $\rm SO_2$ lines show the emission local maximum within B2. 

All of the SO and $\rm SO_2$ lines show the strongest emission toward B1a. More specifically, $\rm ^{34}SO$, $\rm SO_2$, and $\rm ^{34}SO_2$ lines  show the U-shape structure toward the wall and southern rim of B1, linking B0e-B1a-{  B1c}-{  B1b}.  In contrast, SO lines suggest an anticorrelated spatial distributions, having a bright emission toward the eastern B1a.   This finding could either reflect a chemical segregation or be due to optical depth effects, and it will be investigated in the next sections.

\begin{table}
\small
\caption{Positions corresponding to the clumpy structures shown on the maps in  Figure~\ref{1157:smalldistribution}.}\label{tab:clump}
\centering
 \scalebox{1}{
\begin{tabular}{cp{2.5cm}p{2.3cm}}
\hline
\hline
      &R.A. (J2000) &Dec (J2000)\\

\hline
B0e          &$\rm 20^h39^m10^s.365$     &$\rm 68^{\circ}01\arcmin23.80\arcsec$         \\  
B1a          &$\rm 20^h39^m10^s.301$     &$\rm 68^{\circ}01\arcmin13.51\arcsec$         \\  
{  B1b}          &$\rm 20^h39^m08^s.734$     &$\rm 68^{\circ}01\arcmin10.54\arcsec$         \\  
{  B1c}          &$\rm 20^h39^m09^s.677$     &$\rm 68^{\circ}01\arcmin06.45\arcsec$         \\  
B1f          &$\rm 20^h39^m09^s.444$     &$\rm 68^{\circ}01\arcmin16.70\arcsec$         \\    
{  B1i}          &$\rm 20^h39^m10^s.000$     &$\rm 68^{\circ}01\arcmin01.11\arcsec$         \\  
B2a         &$\rm 20^h39^m12^s.322$     &$\rm 68^{\circ}00\arcmin43.85\arcsec$         \\ 
\hline
\hline

\end{tabular}
}
\end{table}

 \begin{figure*}
\centering
\includegraphics[width=18cm]{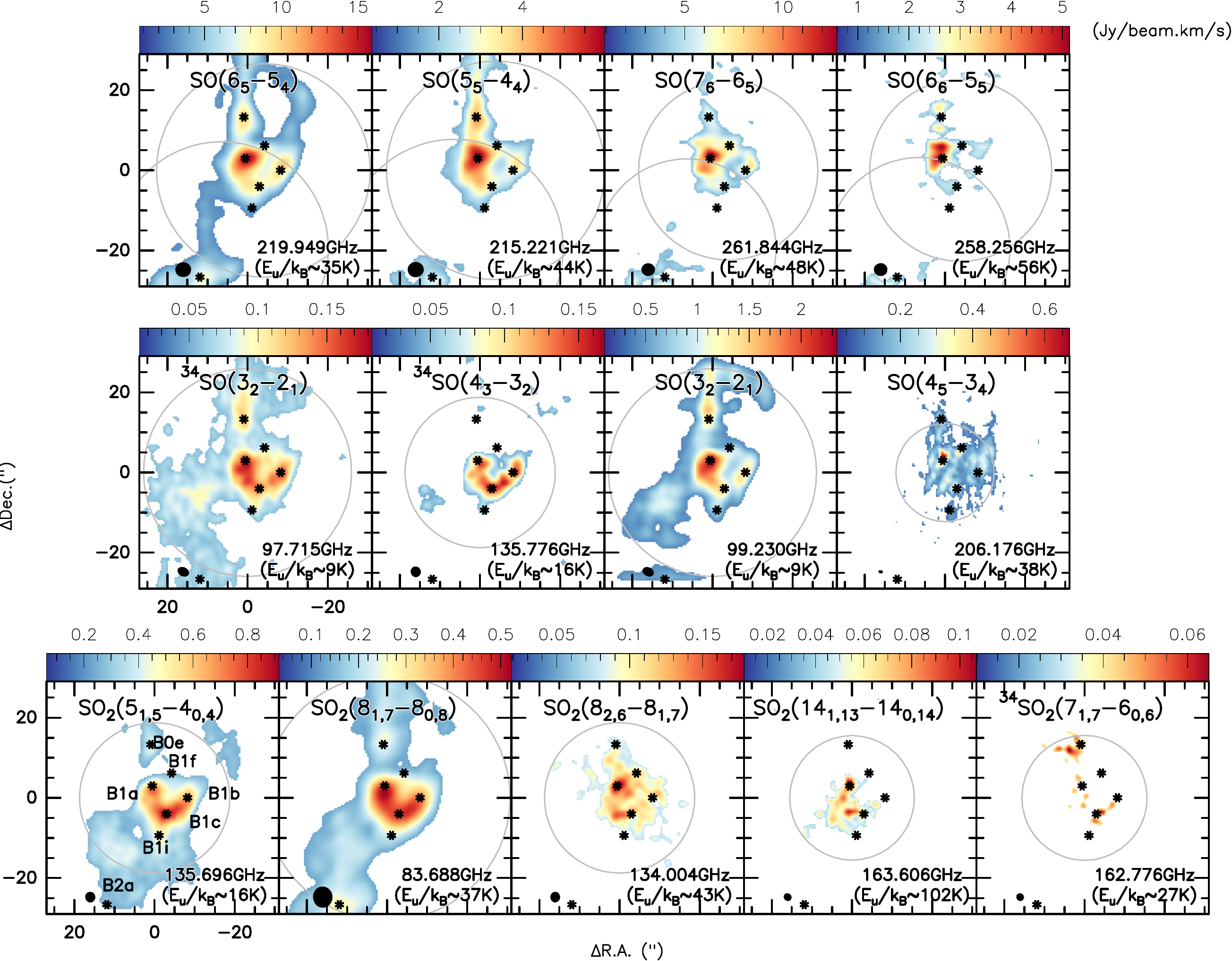}
\caption{
Color maps show the intensity maps of all of the detected $\rm SO_2$ and SO lines integrated over the velocity range of $\rm -10$ to $\rm +5\,km\,s^{-1}$, zoomed in toward the central $\rm 1\arcmin\times1\arcmin$ region. All the maps are obtained from the combination of NOEMA (or SMA) and IRAM-30\,m, and the emission $\rm <5\sigma$ rms are blanked. Seven clumpy substructures labeled as B1a, B1b, B1c, B0e, B1f, B1i, and B2a show strong line emissions.  Gray circle(s) in each panel indicates the primary beam of the NOEMA/SMA at corresponding line rest frequency, and the angular resolution after IRAM-30\,m combination is plotted at the bottom left in each panel.
\label{1157:smalldistribution}}
\end{figure*}

\section{Analysis}\label{1157:analysis}
To quantify the possible chemical differentiations between the SO and $\rm SO_2$ molecules by excluding  excitation effects, we need to  constrain their column densities over the entire targeted region.
 Uncertainty in the molecular column density measurement is mainly from two parts, whether the optical depth of each line can or has to be corrected, and whether these lines can be assumed in  the local thermal equilibrium (LTE) condition.

\subsection{Optical depths}\label{sec:34S32S}
In the shocked region, we cannot exclude the possibility that some low-$J$ SO and $\rm SO_2$ lines may be optically thick. 
We note that  $\rm ^{34}SO\, (3_2-2_1)$ and $\rm ^{32}SO\, (3_2-2_1)$ are a pair of lines that have similar  upper-level energies ($E_u/k_B\sim9$\,K) and  Einstein coefficients ($A_{ij}\sim\rm 10^{-6}\,s^{-1}$). Therefore, their brightness temperature ratio, $\rm T_{B, ^{34}SO{(3_2-2_1)}}/T_{B, ^{32}SO{(3_2-2_1)}}$, can be used to measure the optical depth $\tau$  of  $\rm ^{32}SO\, (3_2-2_1)$ line \citep{myers83}\footnote{Assuming that fractionation (isotopic exchange reaction) of $\rm ^{34}S\Leftrightarrow {^{32}S}$ is stable, according to \citet{myers83}, we can estimate the optical depth at the line center of a given line $\rm \tau_{\alpha,0}$ by measuring the ratio between the main beam brightness temperature of the main line $\rm T_{mb,~\alpha,0}$ and its  isotopologue $\rm T_{mb,~\beta,0}$ as $\rm \frac{1- exp(-\tau_{\alpha,0}/\mathcal{R}_{\alpha})}{1-exp(-\tau_{\alpha,0})}\approx \frac{T_{mb, ~\beta,0}}{T_{mb, ~\alpha,0}}$. Here, $\mathcal{R}_{\alpha}$ is the intrinsic abundance of the main isotope (e.g., $\rm ^{32}S$) compared to its rare  isotope (e.g., $\rm ^{34}S$) in the ISM. 
A numerical solver to this approximation indicates that  $\tau_{\alpha,0}$  is  appropriate to $\sqrt{\rm ln(\frac{T_{mb, ~\beta,0}}{T_{mb, ~\alpha,0}})}$  at each pixel.
}. 

Figure~\ref{1157:34S32S} shows that such ratio is in the range of 0.1--0.15 toward the pixels where both lines show $\rm>5\sigma$ emissions. Considering the 10\% systematic uncertainty per pixel, $\rm T_{B, ^{32}SO{(3_2-2_1)}}/T_{B, ^{34}SO{(3_2-2_1)}}$, toward pixels within B0-B1 can be treated as uniform. Therefore, if we assume that $\rm ^{34}SO{(3_2-2_1)}$ is optically thin, we can reasonably assume that $\rm ^{32}SO{(3_2-2_1)}$ has uniform optical depths toward all valid pixels over the entire B0-B1 region. 

Quantitatively, using the standard isotopic ratio  $\mathcal{R}_{\rm ^{32}S/^{34}S}\sim22$  in the solar system \citep[e.g.,][]{wilson94,chin96,lodders03},  $\tau$ of the $\rm ^{32}SO\,{(3_2-2_1)}$ line  in B0-B1 can be estimated. 
Solving numerically the \citet{myers83} approximation (see footnote 7),
we present the  range of $\tau$ on Figure~\ref{1157:34S32S} as well. 
In general,  $\tau$ of the $\rm ^{32}SO\,{(3_2-2_1)}$  line is not negligible ($\rm \leq 1.5$ toward B0e and B1a, as well as $\rm \leq 3$  toward B1f, B1c, and B1i, and $\rm \leq 4$ toward the western edge), but it has small dynamic range in B0-B1 region.
As the absolute value of  $\tau$ is an approximation and depends on the assumption of $\mathcal{R}_{\rm ^{32}S/^{34}S}$, $\tau$ toward individual pixels is not important in this work. Instead, the relative gradient of $\tau$ from B0e to B1a and B1i is $\sim$1, indicating that the relative gradient of SO column density in B0-B1 does not change significantly before and after line optical depth correction.

We are not able to estimate the $\tau$ of any $\rm ^{32}SO_2$ line, due to the lack of detections on their $\rm ^{34}SO_2$ line counterparts (with the same transition level). 
Nevertheless,  the relative abundance ratio between this isotopologue pair is comparable with the standard Solar $\mathcal{R}_{\rm ^{32}S/^{34}S}$  (see Section \ref{sec:trot}), which  indicates that the optical depths of the $\rm SO_2$ lines may not lead to an underestimation of the $\rm SO_2$ column density  in this work.

  \begin{figure}
\centering
\begin{tabular}{cc}
\includegraphics[height=7.6cm]{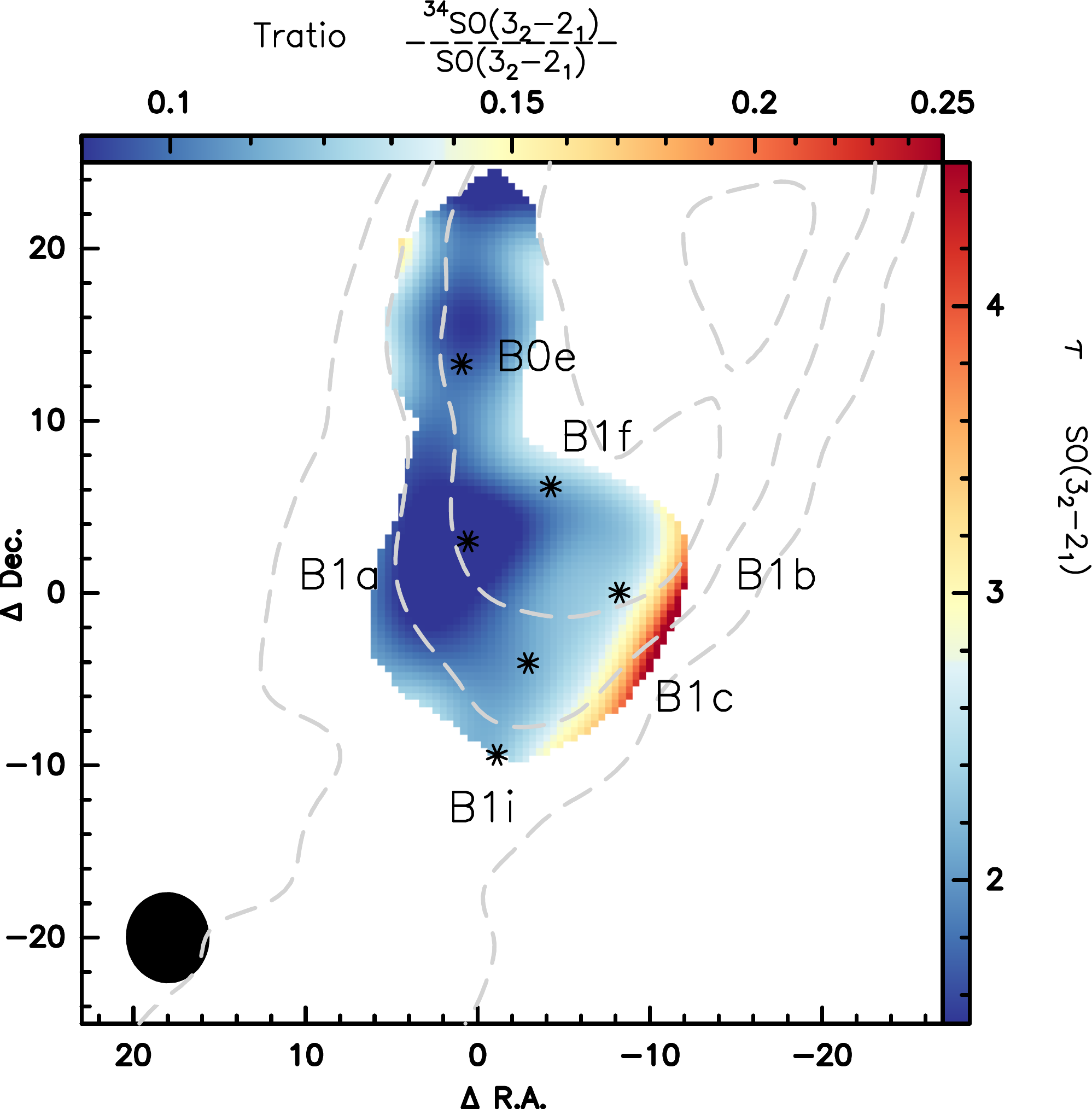}
\end{tabular}
\caption{Map of the integrated intensity  ratio $\rm T_{B, ^{34}SO_{(3_2-2_1)}}/T_{B, ^{32}SO_{(3_2-2_1)}}$  (top widget)  as well as the optical depth of  $\rm ^{32}SO\,{(3_2-2_1)}$ (right widget)  toward each pixel. 
The gray dashed contours, starting from $\rm 8\sigma$ ($\rm \sigma=0.27\,Jy\,beam^{-1}$) and increasing with the step of  $\rm 8\sigma$, show the CO (1-0) emission \citep{gueth96}. The pixels where $\rm ^{34}SO\,{(3_2-2_1)}$ shows $\rm <6\sigma$ emissions are blanked. Six clumpy structures toward B1 are labeled. 
\label{1157:34S32S}}
\vspace{0.8cm}
\end{figure}

\subsection{Rotation temperature and molecular column density from Rotational Diagram (RD)}\label{sec:trot}
Although the line emissions are clumpy, their spatial distributions toward the positions list in Table~\ref{tab:clump} are more extended than their synthesized beams, so we do not take the beam filling factor into account in the following estimates. As the first-order approximation, we assume that all of the detected SO and $\rm SO_2$ lines are optically thin and under LTE. Smoothing the line-intensity maps to the same angular resolution (i.e., $\rm 5.30\arcsec\times4.86\arcsec$) and the same pixel size  (i.e., $\rm 0.5\arcsec$), we use the RD method, and fit  the relative population distribution of all the  detected energy levels by assuming a single component toward each clumpy substructure (see RD fit toward B1a in Figure~\ref{1157:rotdiaSOSO2} as an example).  The least-square solutions of the molecular rotational temperature $\rm T_{rot}$, as well as total column density  $\rm N_{rot}$ from fittings are list Table~\ref{rdlvgfit}.

When taking the optical depth of the $\rm SO\,{(3_2-2_1)}$ lines into account, the SO column density in pixels within B0-B1  increases by a factor of 1.5--2.5  (Table~\ref{rdlvgfit}). 
{\color{black}
Taking into account the uncertainties, it seems that the difference between the SO column density throughout the B0-B1 structure is little, ranging from (3.9 $\pm$ 1.7) $\times$ 10$^{14}$ cm$^{-2}$ (B1a) to (1.5 $\pm$ 1.0) $\times$ 10$^{14}$ cm$^{-2}$ ({  B1i}).}
Also, the SO$_2$ column densities, as derived from RD, looks quite uniform, being around 1 $\times$ 10$^{14}$ cm$^{-2}$.
SO seems to have uniform $\rm T_{rot}$ toward the B0-B1 region (10--14\,K), and the optical depth correction for  the $\rm SO\,{(3_2-2_1)}$ line does not bring in significant change in the $\rm T_{rot}$ estimates. In contrast, precise $\rm T_{rot}$  estimation of $\rm SO_2$ depends on whether the high-$J$ line (with $E_u/k_B>$100\,K) is included, i.e., fitting to the mid-$J$ $\rm SO_2$  lines  yields the same $\rm T_{rot}$ as that from fitting the low-$J$ and mid-$J$ SO lines  (with  $E_u/k_B<$70\,K). A possible reason could be that the high-$J$ $\rm SO_2$ may trace a warmer component.

 \begin{figure*}
\centering
\begin{tabular}{cc}
\includegraphics[height=4.5cm]{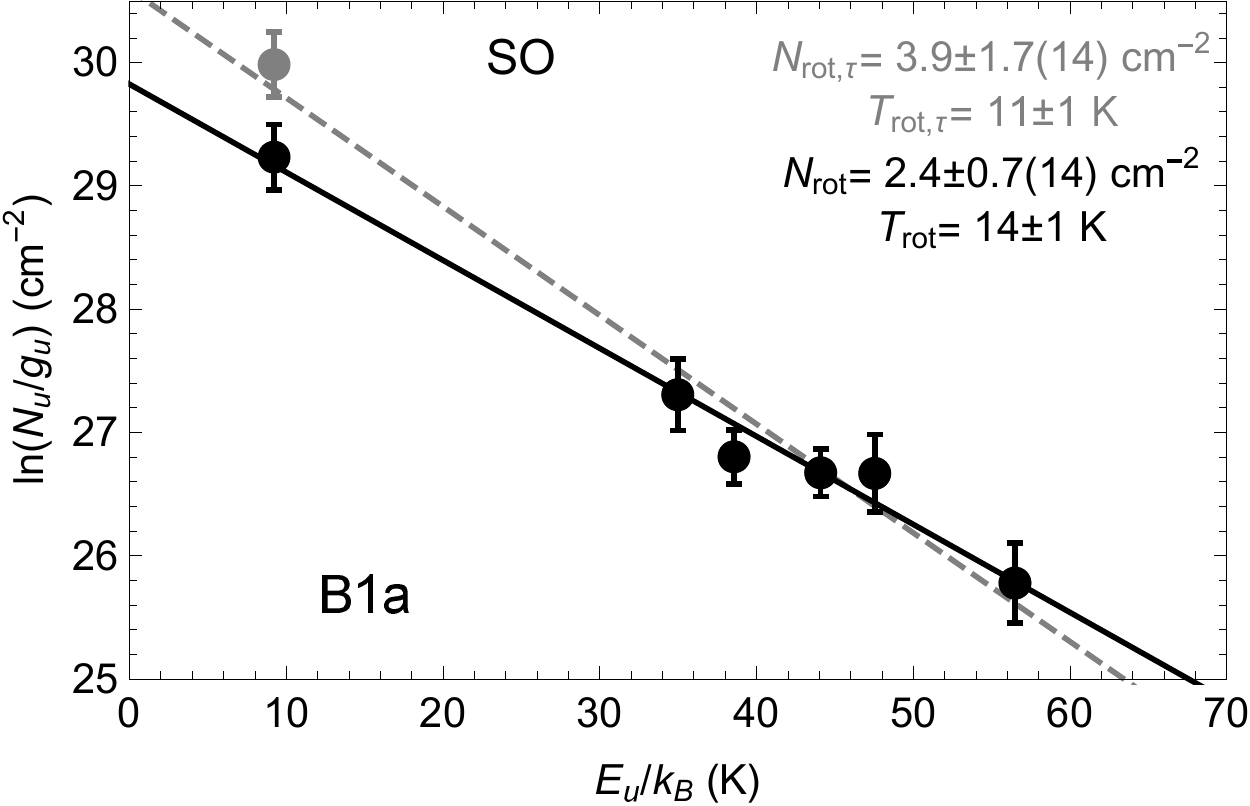}
\includegraphics[height=4.5cm]{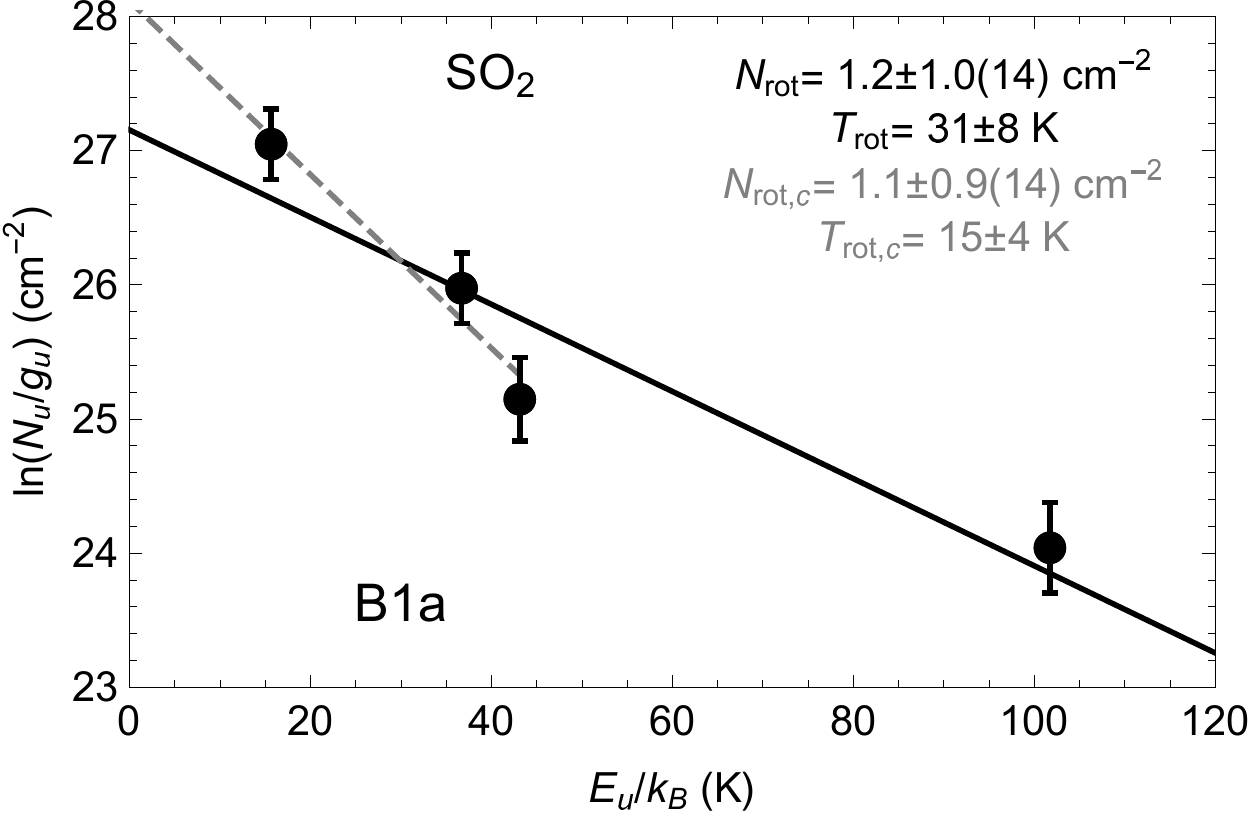}

\end{tabular}
\caption{Rotation diagrams of SO ($left$) and $\rm SO_2$  ($right$)  toward B1a as an example. The black dots correspond to the line emission integrated in the velocity range of $\rm -10$ to $\rm +5\,km\,s^{-1}$. 
The error bar of each data point comes from the uncertainty in observations and data reduction (calibration, combination, and intensity integration). The black line is fitted with the assumption that lines from the same species are optically thin, under LTE, and trace a single component toward B1a. 
The gray dashed  line and gray dot in the $left$ panel is fitted by taking the optical depth of $\rm SO\,{(3_2-2_1)}$ into account.
The gray dashed  line in the $right$ panel is fitted by only taking the low-$J$ and mid-$J$ lines with $E_u/k_B<70$\,K into account.
\label{1157:rotdiaSOSO2}}
\end{figure*}

The upper energy level of the detected $\rm ^{34}SO$ lines are too close to derive the molecular rotational temperature map, and we are not able to obtain the rotational temperature of $\rm ^{34}SO_2$ from only one detected line $\rm (7_{1,7}-6_{0,6})$. 
With the LTE and optically thin assumptions, we estimate the  column densities of $\rm ^{34}SO$  and $\rm ^{34}SO_2$ by using the $\rm T_{rot}$ of their $\rm ^{32}S$- isotopologues  toward the clumpy substructures in B0-B1.  The relative abundance ratio between each pair of  $\rm ^{32}S$- and $\rm ^{34}S$- isotopologues  is uniform (with a small standard deviation among pixels), especially toward B1.  The ratio of $\rm \chi({^{32}SO/^{34}SO})$ is $\rm 9\pm2$ over the entire B0-B1  ($\sim7$ toward B0e and $\rm 15\pm4$ toward B1 after $\rm SO\,{(3_2-2_1)}$ optical depth correction),  while the ratio of  $\rm \chi({^{32}SO_2/^{34}SO_2})$ is $\rm \sim6$ toward B0e and $\sim 15$ toward B1a-B1c-B1b (assuming $ \tau\rm \le1$ of all of the $\rm SO_2$ lines detected in this work). These ratios are consistent with  the $\rm \chi({C^{32}S/C^{34}S})\sim 15$ measured toward B1 from our previous single-dish observations \citep{bachiller97}\footnote{The column density of $\rm C^{34}S$ given in \citet{bachiller97} was a typo, and it is corrected as ``1.8(13)".}, and is comparable with the $\rm \mathcal{R}_{^{32}S/^{34}S}\sim 22$ in the solar system. Therefore, the optically thin assumption for the $\rm SO_2$ lines can be used to measure the $\rm SO_2$ column density  in this work.

\subsection{Kinetic temperature and molecular column density from  Large Velocity Gradient (LVG)}\label{sec:lvg}

In the temperature range of 10--100\,K, the critical densitie of all of the detected SO and $\rm SO_2$ isotopologue lines are in the range of  $\rm 10^5-10^6\,cm^{-3}$ (estimated by using the Einstein coefficient and the collision rate given in Leiden Atomic and Molecular Database, \citealp{schoier05}, which are consistent with the effective critical density given by, e.g., \citealp{reiter11}). 
According to previous studies, such as \citet[][]{benedettini13,gomezruiz15}, the number density toward the entire  B1 region is $\rm 10^5\text{--}10^6\,cm^{-3}$. Therefore, excitation condition of SO and $\rm SO_2$ gas in this shocked region need a LVG consideration. 

Using the  collision coefficients of SO-(para) $\rm H_2$ and $\rm SO_2$-(para and ortho) $\rm H_2$ obtained from the  BASECOL database \citep{dubernet13}, and applying the code of LVG approximation to non-LTE analysis described in \citet{ceccarelli02}, we run grids of parameters  and find the least-square ($\rm \chi^2$) best-fitting solution toward four positions along the precessing path, i.e., B0e, B1a (as well as {  B1b} as a double-check), and {  B1i}. B2a is not fit because less than four lines show $\rm >5\sigma$ emission.
In the fittings, we adopted a FWHM linewidth of $\rm 7\,km\,s^{-1}$  toward each position.
Then, we vary three parameters in a relatively large range: $\rm H_2$ number density $n$ ($\rm 10^4\,cm^{-3}$--$\rm 10^8\,cm^{-3}$), gas kinetic temperature $\rm T_{kin}$ (30\,K--300\,K), molecular column
density $\rm N_T$ ($\rm 10^{13}\,cm^{-2}$--$\rm 10^{18}\,cm^{-2}$). 
Note that the FWHM linewidth we adopt in this work is a compromise made to obtain the best fit by including all SO and $\rm SO_2$ lines observed at a coarse velocity resolution (6--7\,$\rm km\,s^{-1}$). According to the previous analysis based on CO observations at a velocity resolution of 0.1--0.5\,$\rm km\,s^{-1}$ \citep{lefloch12}, at least three components overlap in the velocity range from -30 to 5\,$\rm km\,s^{-1}$, with their intensity peaks close to the $\rm V_{sys}$. Before we resolve them with observations at higher-velocity resolution, the adopted FWHM linewidth in the range of 5--10\,$\rm km\,s^{-1}$
provides results that represent the average behavior of the different kinematical components within this velocity range.

Comparing the best-fit results for the column density by using LVG and RD methods (Table~\ref{rdlvgfit}), we have
a good agreement. In addition, we find that the SO column density toward B1i is slightly lower than those measured toward the rest
of the clumps, even after taking into account the uncertainties (see Table~\ref{rdlvgfit}).
The kinetic temperatures are in the range of 30--200\,K, being substantially
in agreement with the kinetic temperatures found by \citet{lefloch12} in the different gas components coexisting in the B1 shocked gas.
On the other hand, the gas volume density $n$ from LVG fit of SO lines is well constrained, from  $\rm (3-5)\times 10^4\,cm^{-3}$ (B0e) to  $\rm 10^5\,cm^{-3}$ ({  B1i}), while  $n$ as traced 
by $\rm SO_2$ ($\rm 10^5\text{--}10^6\,cm^{-3}$) is denser than that traced by SO by a factor up to an order of magnitude (see Table~\ref{rdlvgfit}). The higher values of $n$ are in agreement with those derived by 
\citet{gomezruiz15} using a CS multiline analysis. Therefore, SO and $\rm SO_2$ may be tracing different gas portions.

\begin{table*}
\small
\centering
\caption{Best-fit results by using RD and LVG methods \label{rdlvgfit}}
 \scalebox{1}{
\begin{tabular}{cc|cccc|cccc}
\hline\hline
& &\multicolumn{4}{c|}{RD}    &\multicolumn{4}{c}{LVG}\\       
\hline     
\multirow{2}{*}{Location} &\multirow{2}{*}{Species} &$\rm N_{rot}$   &$\rm T_{rot}$     &$\rm N_{rot,\tau}$$^a$   &$\rm T_{rot,c}$$^b$        &$\rm N_T$                                                   &$\rm T_{kin}$        &$n$     &$\rm \chi^2$   \\
               &                              &($\rm 10^{14}\,cm^{-2}$)     &(K)       &($\rm 10^{14}\,cm^{-2}$)     &(K)        &($\rm 10^{14}\,cm^{-2}$)     &(K)                         &($\rm 10^5\,cm^{-3}$)         &                       \\       \hline

\multirow{2}{*}{B0e}    &SO                     &$\rm 2.1\pm0.8$    &$\rm 13\pm1$    &$\rm 1.8\pm1.3$  &... &$\rm 2.0\pm0.2$   &70--200             &0.3--0.5        &0.94\\  
                                   &$\rm SO_2$       &$\rm 1.0\pm0.8$    &$\rm 28\pm9$   &...  &$\rm 12\pm3$ &$\rm --$$^c$    &$\rm --$$^c$           &$\rm --$$^c$         &$\rm --$$^c$ \\  
\hline
\multirow{2}{*}{B1a}    &SO                     &$\rm 2.4\pm0.7$    &$\rm 14\pm1$    &$\rm 3.9\pm1.7$  &...  &$\rm 2.5\pm0.5$   &40--70             &0.8--0.9        &0.6\\  
                                   &$\rm SO_2$       &$\rm 1.2\pm1.0$    &$\rm 31\pm8$   &... &$\rm 15\pm4$   &$\rm 1.4\pm0.4$   &$\rm 50\pm10$             &$\rm 3\pm2$      &0.5\\     
                                   
\hline
\multirow{2}{*}{{  B1b}}    &SO                     &$\rm 2.3\pm1.4$    &$\rm 13\pm1$  &$\rm 3.5\pm2.0$   &...  &$\rm 2.0\pm0.2$   &30--100             &0.8--1        &1.2\\  
                                   &$\rm SO_2$       &$\rm 1.2\pm1.0$    &$\rm 25\pm5$  &...  &$\rm 15\pm4$   &$\rm 1.0\pm0.2$   &$\rm 30\pm20$             &$\rm 10\pm5$      &0.7\\          
                                   
\hline
\multirow{2}{*}{{  B1i}}    &SO                    &$\rm 1.4\pm0.8$    &$\rm 13\pm1$    &$\rm 1.5\pm1.0$   &...   &$\rm 1.4\pm0.2$   &30--100             &0.8--1        &0.5\\  
                                   &$\rm SO_2$      &$\rm 1.4\pm0.7$    &$\rm 28\pm9$  &... &$\rm 12\pm3$  &$\rm 1.2\pm0.2$   &$\rm 50\pm5$             &$\rm 3\pm1$      &0.5\\

\hline
\multirow{2}{*}{B2a}    &SO                    &$\rm 0.8\pm0.5$    &$\rm 19\pm3$    &... &...  &$\rm --$$^c$    &$\rm --$$^c$            &$\rm --$$^c$        &$\rm --$$^c$\\  
                                   &$\rm SO_2$     &$\rm<1.7$$^d$     &...   &...&$\rm< 64$$^d$    &$\rm --$$^c$    &$\rm --$$^c$             &$\rm --$$^c$       &$\rm --$$^c$\\   
\hline
\hline
\multicolumn{10}{l}{{\bf Note.} {\it a}. Column density estimated by taking the optical depth of $\rm SO\,{(3_2-2_1)}$ into account. }\\
\multicolumn{10}{l}{~~~~~~~~~~~{\it b}.  Rotation temperature  derived by only taking the lines with $E_u/k_B\rm <70$\,K into account.}\\ 
\multicolumn{10}{l}{~~~~~~~~~~~{\it c}.  Less than four lines show emission with $\rm S/N>5$ in the primary beam.}\\ 
\multicolumn{10}{l}{~~~~~~~~~~~{\it d}.  Upper limit is given by using two lines which show emission with $\rm S/N>5$ .}\\ 
\end{tabular}

}

\end{table*}

\section{Discussion}\label{1157:discussion}

\subsection{Map of relative abundance ratio between SO and $\rm SO_2$}\label{1157:ratio}

Provided that the column densities as derived by the LVG analysis are in agreement
with those derived from the LTE-RD analysis (after correction due to the
optical depths), to investigate the possible chemical differentiation
between SO and $\rm SO_2$ along the path of the precessing jet in the
entire B0-B1-B2 (in a time frame of thousand years), we fit the
one-component RD toward pixels in B0-B1-B2 region.

When investigating
the 0.1\,pc scale field of view, we note that there is a tentative
gradient along the path of the precessing jet in the
$\rm \chi({SO/SO_2})$ map before optical depth correction\footnote{\color{black}Because of the line image sensitivity,  the optical depth correction only works for the $\rm SO\,{(3_2-2_1)}$ line toward the B0-B1 region.  As a compromise for inspecting the map of column density ratio between SO and $\rm SO_2$ in a larger field, we assume that  all lines are optically thin and under LTE in Figure~\ref{1157:rotmap}.}  (Figure~\ref{1157:rotmap}):
$\rm \chi({SO/SO_2})<1$ in B2, while $\rm 1.5<\chi({SO/SO_2})<3$ from the younger B0e. 
 Unfortunately,  only two $\rm SO_2$ lines show $\rm >5\sigma$ emission toward B2, so $\rm SO_2$
column density toward B2a is only an upper limit, i.e.,
$\rm \chi({SO/SO_2})$ is a lower limit there. Only further SO$_2$
observations toward B2 will allow us to verify the occurrence of such
a spatial trend.

To test whether the tentative gradient of $\rm \chi({SO/SO_2})$ is due to optical depth effects, 
we zoom into the region where SO column density can be corrected by considering the optical depth of $\rm SO\,{(3_2-2_1)}$ line. 
Even after the correction for optical depth effects, Figure~\ref{1157:ratiomaptau} allows us to speculate a possible trend for the $\rm \chi({SO/SO_2})$ ratio
decreasing moving from north (younger gas) to south (older gas). Again, the present results can only suggest the
occurrence of such a spatial trend and only future multiline observations, possibly on smaller spatial scales, will
verify this speculation.

 \begin{figure*}
\centering
\includegraphics[width=15cm]{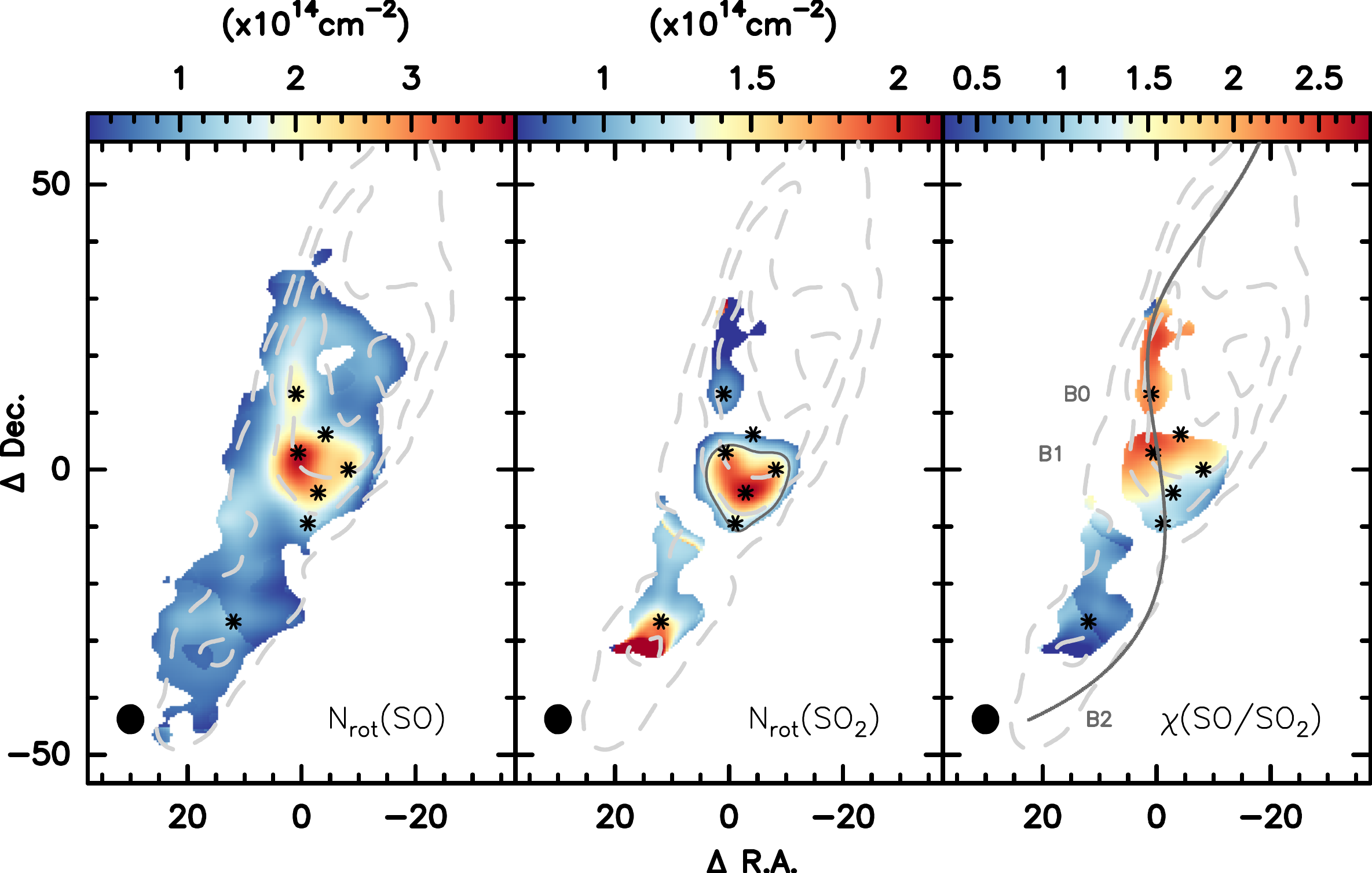}
\caption{SO ({\it left}) and $\rm SO_2$ ({\it middle}) column density  maps and their relative abundance ratio map ({\it right})  derived by using the RD method, with the assumption that all lines are optically thin and under LTE. 
In the {\it left} panel, the pixels where less than three SO lines show $\rm >5\sigma$ emission are blanked.  
In the {\it middle} panel, the pixels where less than two $\rm SO_2$ lines show $\rm >5\sigma$ emission are blanked, and the black contour denotes the region within which at least three lines show $\rm >5\sigma$ emission.  
In the {\it right} panel, the pixels where less than two $\rm SO_2$ lines show $\rm >5\sigma$ emission are blanked, and the gray line indicates the path of the precessing jet \citep{podio16}.  
These maps are derived by smoothing all lines to the same angular resolution, i.e., $\rm 5.30\arcsec\times4.86\arcsec$, 
shown as the synthesised beam in the bottom left of each panel.
The gray dashed contours and the labeled clumpy structures are the same as shown in Figure~\ref{1157:34S32S}.
\label{1157:rotmap}}
\end{figure*}

\begin{figure}
\includegraphics[angle=0,width=9cm]{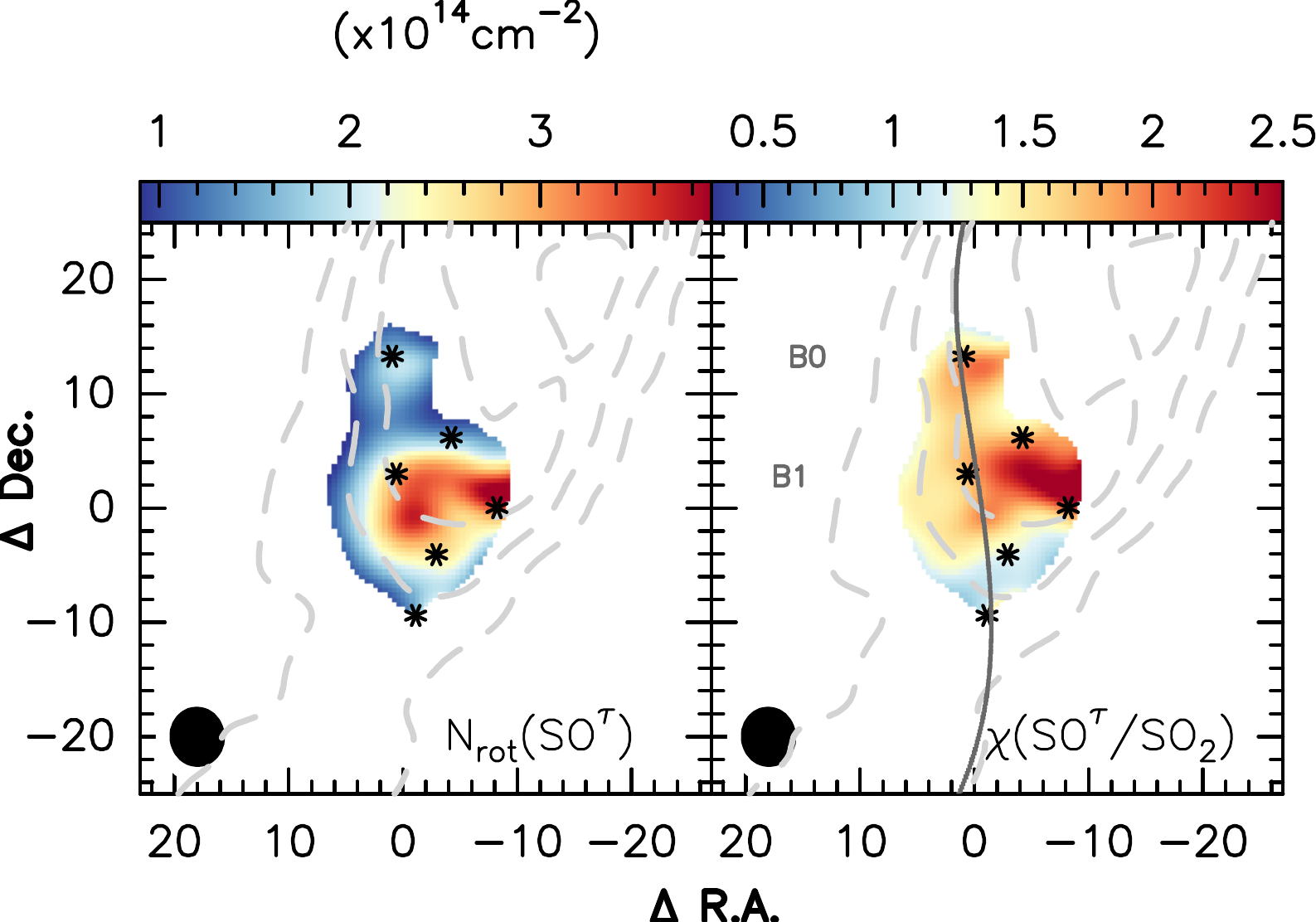}
\caption{SO column density  map ({\it upper}) and map of  its relative abundance ratio with respect to $\rm SO_2$ ({\it lower})  derived by using the RD method, with the LTE assumption and optical depth correction for the  $\rm SO\,{(3_2-2_1)}$ line. 
The pixels where $\rm ^{34}SO\,{(3_2-2_1)}$ shows $\rm <6\sigma$ emissions are blanked.  
In the {\it right} panel, the gray line indicates the path of the precessing jet \citep{podio16}.  
These maps are derived by smoothing all lines to the same angular resolution, i.e., $\rm 5.30\arcsec\times4.86\arcsec$, shown as the synthesised beam in the bottom left of each panel.
The gray dashed contours and the labelled clumpy structures are the same as shown in Figure~\ref{1157:34S32S}.}
\label{1157:ratiomaptau}
\end{figure}

\subsection{Chemical modeling}\label{1157:model}

Our previous chemical study, especially the observations of the ions
and COMs, unveiled at least three chemically
different layers with gas-grain components from the north (B1a) to south
(B1i) toward B1 \citep[][]{codella17,codella20}.
To trace back the possible forming path of SO and $\rm SO_2$
from the $\rm \chi({SO/SO_2})$ gradient through B0-B1-B2, we apply the
same time-dependent gas-phase chemical model as that introduced by
\citet[][]{codella17}. Briefly, we used the MyNahoon code, which is
based on the Nahoon code originally developed by \citet{wakelam12}. 
The code follows the gaseous abundances evolution in time.
Note that it does not consider the reactions possibly occurring on the grain
surfaces, except the formation of H$_2$ molecules, as our goal is to
study a subsequent phase: namely, when the grain mantles are already
formed, in conditions and timescales where the surface reactions are
completely negligible (see later).
In short, we simulate the physical effects caused by the shock passage
by suddenly increasing the H$_2$ gas number density $n$ from $10^4$ to
$\rm 10^5$ cm$^{-3}$, and the temperature from 10 to 70 K.

In addition, because of the sputtering and/or grain-grain collisions
caused by the shock, a number of species previously frozen on the grain mantles
are injected into the gas phase. In our model, we assumed the injected
species and their abundance as those in \citealp{codella17} (their
Table B.1), which is based on several previous studies of the same
source \citep{tafalla95,benedettini13,podio14,codella15} or
known properties of the ice mantle compositions
\citep{boogert15,busquet13}.

In particular, with respect to the S-bearing frozen species,
\citet{podio14} could only roughly constrain them, based on the
S-bearing ions SO$^+$ and HCS$^+$. These authors found that OCS is
injected in the gas phase with an abundance of $6\times10^{-6}$ (with
respect to the H nuclei) but could not constrain the injected
abundances of $\rm H_2S$ and S, two species that are expected to be also
present in the grain mantles \citep[see e.g., ][]{wakelam04,vidal18}. 
Therefore, we ran models with different injected
abundances of $\rm H_2S$ and S and tried to constrain them based on the
new observations presented in this work.

The results of the modeling are shown in
Figure~\ref{1157:soso2model}. In the figure, we report the model
predicted SO/SO$_2$ abundance ratio as a function of the time, namely
the age of the shocked gas and for different values ($2\times10^{-6}$ and $1\times10^{-8}$) of the injected
$\rm H_2S/H$ and S/H abundances. In the same figure, we also report the values measured
in the previous sections (the uncertainty is given based on RD optically thin assumption, RD
optical correction for the low-$J$ SO line, and LVG)
toward B1a, {  B1b} and {  B1i} and their ages estimated by
\citet{gueth96} and \citet{podio16}. Finally, we also show the model predictions
for values slightly different of the density and cosmic ray ionization
rate to have a rough ``sensitivity'' study and understand how much
the predictions also depend on these, only roughly constrained,
quantities.

The figure shows that the model predictions compare rather well with
the observations and, remarkably, with the increasing ages of B1a ({  B1b})
to {  B1i}. Based on the predictions, the injected $\rm H_2S/H$ and S/H
abundances are equal to $2\times10^{-6}$: injecting pre-frozen OCS
only is not enough to reproduce our new observations, an additional
injection of previously frozen S-bearing species in form of $\rm H_2S$ and
S in approximately the same amount is also needed 
(a similar result was also found by \citealp{holdship16}).
 Second, the cosmic ray ionization rate is around $\rm 3\times 10^{-16}\,s^{-1}$, namely
slightly (a factor 2) lower than that used in previous studies. This
may mean that it slightly decreases going outward from the central
object from which the outflow/jets emanate, which is consistent with
the hypothesis that cosmic rays like particles are accelerated by the
dense shocks at the jet base \citep{padovani16,ivlev19}.

Of course our model, although so far is successful in reproducing
several species (see, for example, \citealt{codella17, codella20}), is rather
simplified; it does not take into account, for example, the
short-lived hot/warm phase of the post-shocked gas, which, in the case
of $\rm SO_2$ could have an impact on the predictions. Having said that,
it is encouraging to see that such a simplified model succeeds to
grab the basic gaseous chemical composition changes  caused by the
shock.

 \begin{figure}
\centering
\includegraphics[width=9cm]{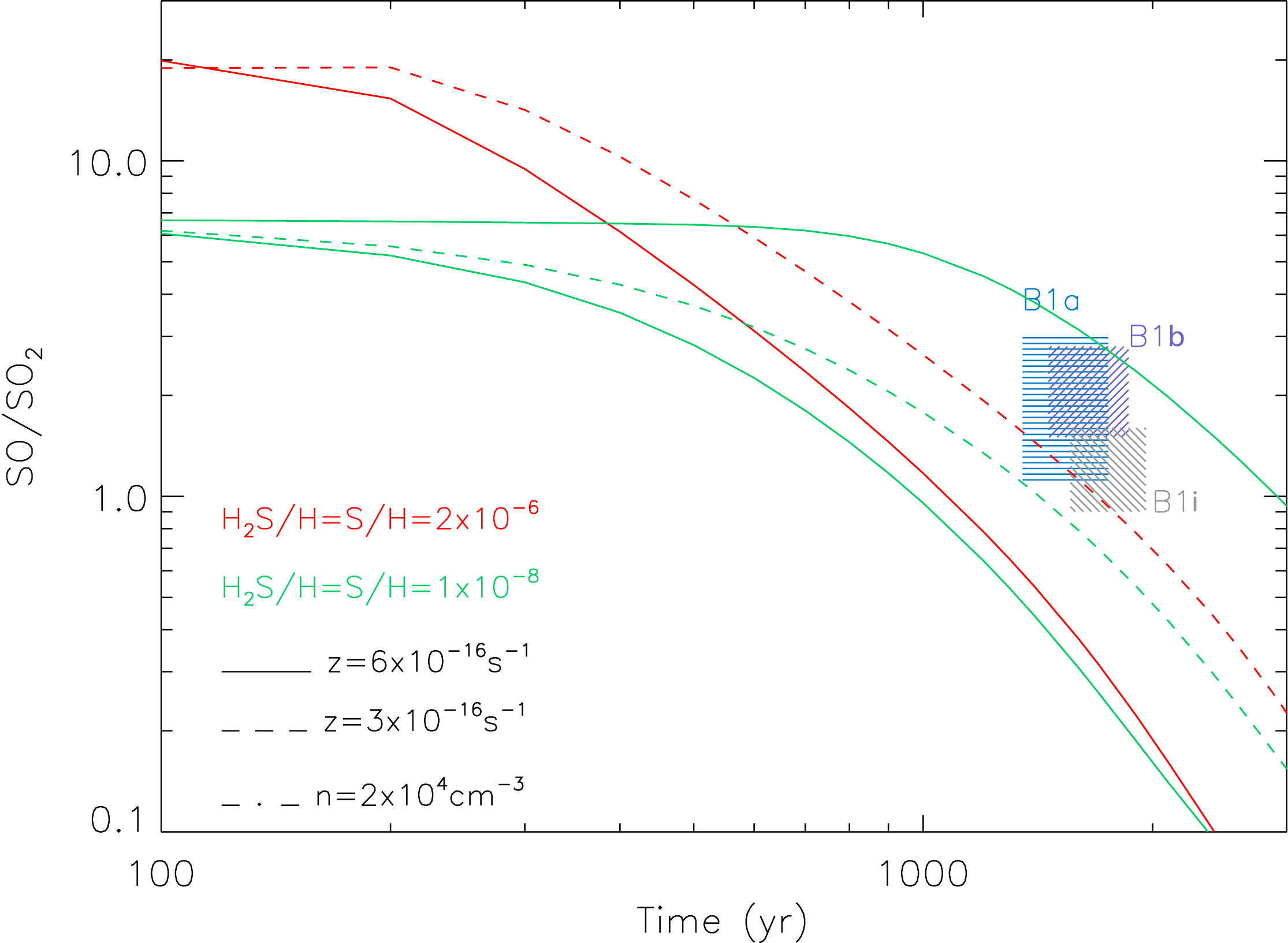}
\caption{SO/SO$_2$ abundance ratio predicted by our astrochemical
  model (see text) as a function of the time for different $\rm H_2S/H$
  and S/H abundances, $2\times10^{-6}$ (red curves) and
  $1\times10^{-8}$ (green curves). The model predictions are also
  obtained for two values of the cosmic rays ionization rates,
  $6\times 10^{-16}$ s$^{-1}$ (solid curves) and  $3\times 10^{-16}$
  s$^{-1}$ (dashed curves), and a smaller H density, $2\times10^4$
  cm$^{-3}$ (dotted-dashed curve). The results from the analysis of
  the observations are reported in the dashed areas toward three
  positions along the B1 region: B1a,  B1b, and B1i, respectively. }
\label{1157:soso2model}
\end{figure}

\section{Conclusions}\label{1157:conclusion}

We simultaneously analyze the chemical properties of SO and $\rm SO_2$  toward all of the brightest shocked regions along the L1157 chemically rich blueshifted outflow, namely, B0, B1, and B2. Based on the data taken in the context of the SOLIS IRAM-NOEMA Large Program, and
the supplementary observations from SMA interferometer as well as IRAM-30\,m telescope at 1.1--3.6\,mm wavelengths, our conclusions are as follows:\\

\begin{enumerate}
\item At an angular resolution of 1\arcsec--3\arcsec~(500--1400\,au at the source distance of 352\,pc), 
SO and $\rm SO_2$ may trace different gas toward B1. This is indicated from the different  
spatial distributions of five $\rm SO_2$ lines (including one $\rm ^{34}SO_2$ line) and eight $\rm SO$ lines (including two $\rm ^{34}SO$ lines) which we detected.
The emission peak of the SO lines are in general toward northeast, 
where the gas is impacted by the recent jet impinging on the B1 cavity wall;  while the emission peak of the $\rm SO_2$ lines are all toward the southern edge of the  older shock front. 
Moreover,  LVG constraints on gas volume density indicates that, the gas traced by $\rm SO_2$  has a volume density of $\rm 10^5\text{--}10^6\,cm^{-3}$, which  is denser than that traced by SO by a factor up to an order of magnitude.

\item Investigating the 0.1\,pc scale field of view, there is a tentative  gradient in the relative abundance ratio of $\rm \chi({SO/SO_2})$ along the path of the associated precessing jet, showing smooth decrease from B0-B1a ($\rm >2$) to B1i-B2 ($\rm <1$).  
These findings call for further analysis at higher spatial resolutions to verify the tentative trends.

\item Our astrochemical modeling shows that the SO and $\rm SO_2$ abundances
evolve on timescales less than about 1000\,yr.
The computed values of timescales toward individual pixels are consistent with the ages estimated in L1157 B0-B1 by previous studies. 
Furthermore, the modeling requires high abundances of both $\rm H_2S/H$ and S/H
($2\times10^{-6}$) injected in the gas phase due to the shock occurrence,
so pre-frozen OCS only is not enough to reproduce our new observations.

\end{enumerate}

\begin{acknowledgements}
We thank the NOEMA, SMA, and IRAM-30\,m staff for
their helpful support during the observations and data reduction of the {\it SOLIS} project. 
We also thank the referee for constructive suggestions.

S.F. acknowledges the support of National Natural Science Foundation of China No. 11988101, the  support of  the CAS International Partnership Program No.114A11KYSB20160008, and the support of the EACOA fellowship from the East Asia Core Observatories Association (EACOA).  EACOA consists of the National Astronomical Observatory of China, the National Astronomical Observatory of Japan, the Academia Sinica Institute of Astronomy and Astrophysics, and the Korea Astronomy and Space Science Institute. 

This work was also supported by (i) the French program ``Physique et Chimie du Milieu Interstellaire" (PCMI) funded by the Conseil National de la Recherche Scientifique (CNRS) and Centre National d'Etudes Spatiales (CNES), (ii) by the Italian project PRIN-INAF 2016  The Cradle of Life - GENESIS-SKA (General Conditions in Early Planetary Systems for the rise of life with SKA), (iii) by the program PRIN-MIUR 2015 STARS in the CAOS - Simulation Tools for Astrochemical Reactivity and Spectroscopy in the Cyberinfrastructure for Astrochemical Organic Species (2015F59J3R, MIUR Ministero dell'Istruzione, dell'Universit\`a della Ricerca e della Scuola Normale Superiore), (iv) by the French Agence Nationale de la Recherche (ANR), under reference ANR-12-JS05-0005, and (v) by the European Research Council (ERC) under the European Union's Horizon 2020 research and innovation programme, for the Project ``The Dawn of Organic Chemistry" (DOC), grant agreement No. 741002, (vi) the Ministry of Science and Technology (MoST) of Taiwan (grant No. 108-2112-M-001-002-MY3).
This research made use of NASA's Astrophysics Data System.
 \end{acknowledgements}

\software{GILDAS/CLASS \citep{pety05}, MIR \citep{scoville93}, MIRIAD \citep{sault95}}

\bibliographystyle{aa}
\bibliography{SO2_ll157-arxiv.bbl}

\newpage
\setcounter{section}{0}
\renewcommand{\thetable}{A\arabic{section}}
\setcounter{table}{0}
\renewcommand{\thetable}{A\arabic{table}}
\setcounter{figure}{0}
\renewcommand{\thefigure}{A\arabic{figure}}
\appendix


Figure~\ref{1157:velpro} shows the profile of the SO and $\rm SO_2$ isotopologue lines we use to measure the molecular column densities toward seven positions.

Table~\ref{NOEMAconfI}  lists the NOEMA observation dates, spectral setup, configurations, calibrators, and weather conditions.

\begin{figure*}
\begin{center}
\includegraphics[angle=90,scale=0.85]{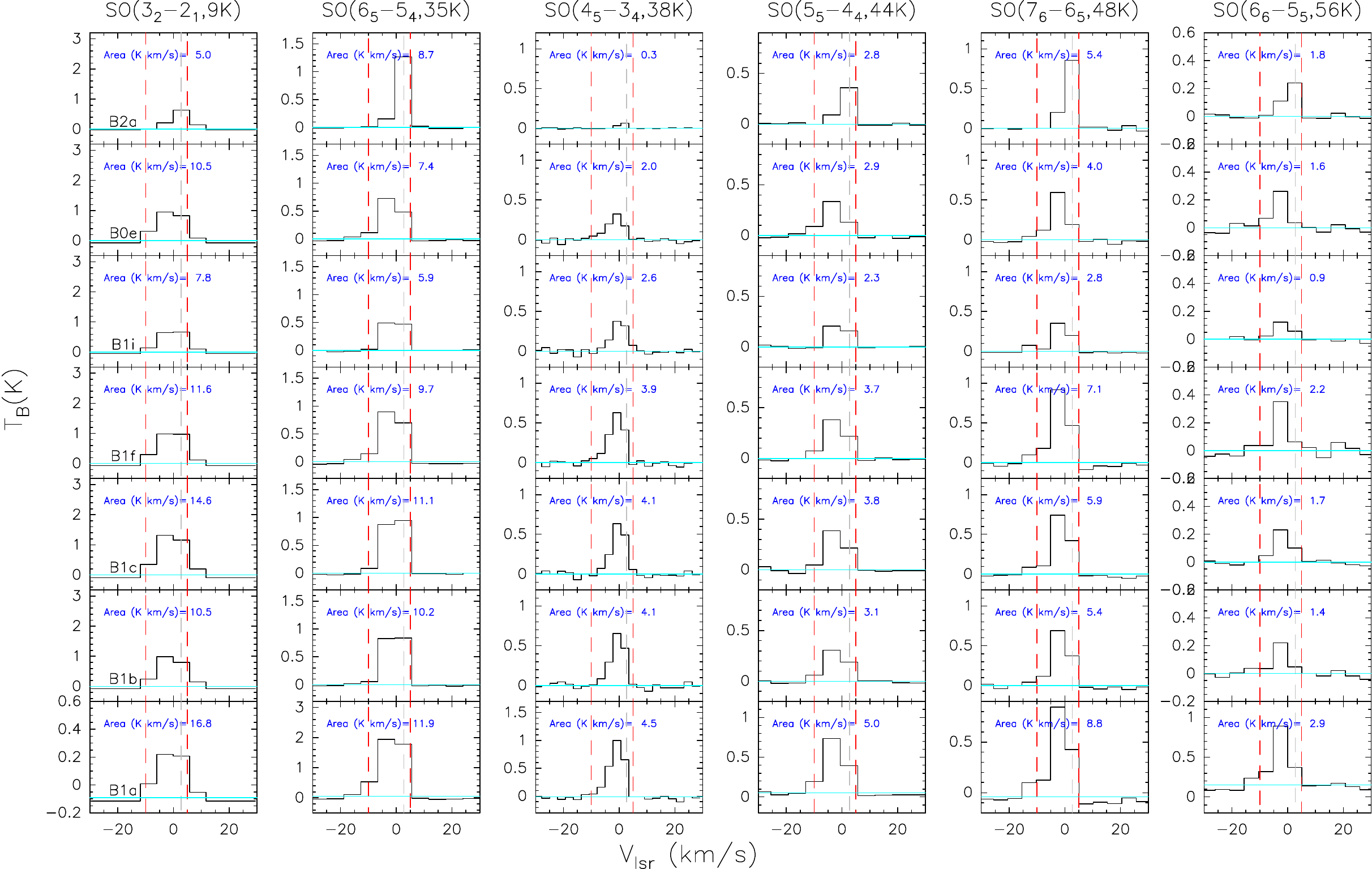}
\end{center}
\caption{Profiles of identified SO and $\rm SO_2$ isotopologue lines, averaged from a beam-sized region with the center  toward each clumpy substructure in the plane of the sky  (Table~ \ref{tab:clump}). All lines are extracted from images smoothed to the same pixel size and angular resolution ($\rm 5.1\arcsec$).  Their velocity resolutions are listed in Table~\ref{line}. In each panel, two red vertical lines at $\rm -10\,km\,s^{-1}$ and $\rm +5\,km\,s^{-1}$ indicate the velocity range for which we integrate the intensity; the gray dashed vertical line indicate the $\rm V_{sys}=2.7\,km\,s^{-1} $ of the cloud. The horizontal cyan line indicates the baseline ($\rm T_B$=0\,K).
}\label{1157:velpro}
\end{figure*} 

\setcounter{figure}{0}
\begin{figure*}
\begin{center}
\includegraphics[angle=90,scale=0.85]{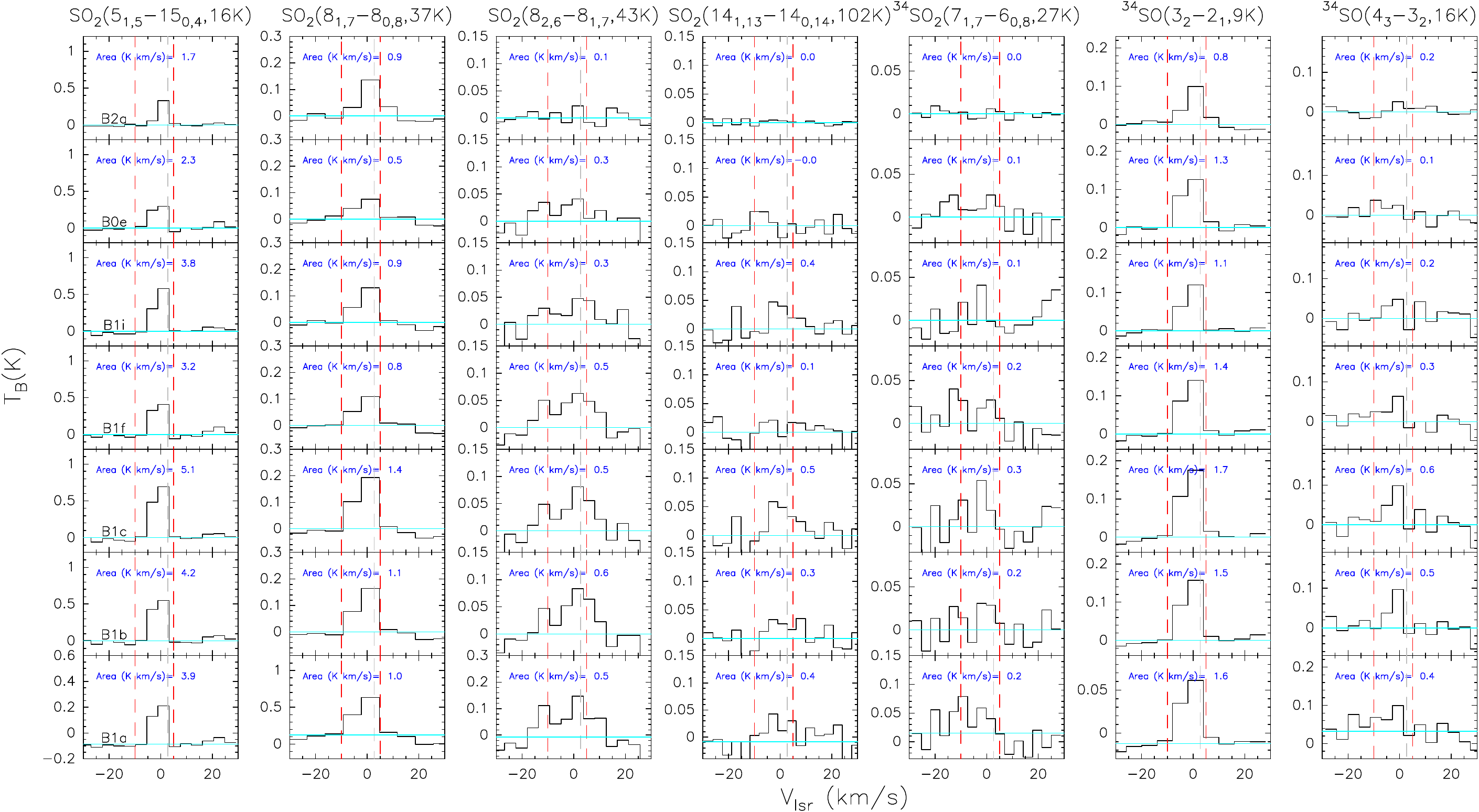}
\end{center}
\caption{(continued)}
\end{figure*}

\newpage

\begin{table*}
\small
\caption{NOEMA observation log toward L1157-B1 \label{NOEMAconfI}}
 \scalebox{1}{
\begin{tabular}{ccp{2.5cm}cp{2.3cm}p{1.5cm}p{1.5cm}p{1.5cm}c}
\hline
\hline
Setup  &WIDEX   &Configuration  &Observation    &\multicolumn{3}{c}{Calibrators}   &$\rm T_{sys}$  &PWV \\
               &(GHz)               &                   &Date                          &(Bandpass)  &(Phase/Amp) &(Flux)      &(K)  &(mm)\\
\hline
1          &80.80-84.40  &D (6ant, 24-97m)        &2015-Jul-15             &1928+738  &1928+738    &MWC349   &100-150  &10-30  \\    
            &                        &                                &2015-Jul-23             &1749+096  &1928+738    &MWC349    &90-160  &10-15 \\      
            &                        &                                &2015-Jul-24             &3C84  &1928+738    &MWC349    &100-300  &20-40 \\    
            &                        &D (5ant, 32-97m)     &2015-Jul-26             &3C454.3  &1928+738    &MWC349    &60-130  &4-8 \\    
            
            &                       &C (7ant, 24-240m)    &2015-Oct-25             &3C454.3  &1928+738    &MWC349    &90-120  &4-6 \\   
            &                       &                                 &2015-Oct-27             &3C84  &1928+738    &LKHA101    &200-400  &10-40 \\    
            &                       &                                 &2015-Nov-04            &3C345  &1928+738    &MWC349    &100-300  &10-40 \\  
            &                       &                                 &2015-Nov-04            &1928+738  &1928+738    &LKHA101    &110-130  &4-8 \\         
\hline

2         &95.85-99.45   &C (7ant, 48-304m)                        &2016-Oct-26                                &0923+392             &2007+659               &MWC349    &100-250  &6-20\\
           &                        &                                                   &2016-Oct-27                                 &1418+546             &2007+659               &MWC349   &50-100  &2-5 \\
           &                        &                                                    &2016-Oct-28                                &2013+370             &2007+659              &MWC349    &50-90  &2-5 \\

           &                     &C (8ant, 24-304m)                        &2016-Oct-22                                  &3C454.3                &2007+659 		&MWC349    &80-250  &4-5\\
           &                     &D (8ant, 24-176m)                        &2017-Jan-22                                  &2200+420             &2007+659           &MWC349   &60-100  &4-5 \\
           &                     &                                                      &2017-Jan-24                                 &2013+370             &2007+659           &MWC349   &50-90  &1-3  \\

\hline
3         &204.00-207.60     &A (7ant, 96-760m)               &2016-Jan-29                                 &2013+370          &1928+738       &MWC349   &200-300  &4-6 \\
           &                         &                                                  &2016-Feb-06                                 &2013+370            &1928+738      &MWC349   &160-300  &3-6  \\  
           &                         &                                                  &2016-Feb-11                                 &3C454.3              &1928+738     &MWC349   &200-400  &3-10\\  
           &                         &                                                  &2016-Feb-14                                 &                            &1928+738      &LKHA101  &130-250  &1-3 \\  
           &                         &                                                  &2016-Feb-16                                 &3C84, 1055+018 &1928+738     &LKHA101  &200-400  &3-4  \\             
           &                      &C (6ant, 24-240m)                        &2016-Apr-20                                 &1949+096           &1928+738      &MWC349    &100-250  &1-4 \\
           &                     &C (7ant, 24-192m)                        &2016-May-06                                &3C454.3              &1928+738    &MWC349   &110-180  &1-2  \\
           &                        &C (7ant, 48-304m)                        &2016-Oct-15                               &3C454.3       &1928+738     &MWC349   &100-150  &2-3 \\

\hline
\hline

\end{tabular}
}
\end{table*}


\end{document}